\documentclass[12pt]{article}

\usepackage{epsf}
\usepackage{amsmath}
\usepackage{graphics}
\usepackage{amsfonts}
\usepackage{amssymb}
\usepackage{latexsym}
\usepackage{color}
\input{colordvi.tex}

\renewcommand{\thefootnote}{\#\arabic{footnote}}

\begin{document}
\setcounter{footnote}{0}

\begin{titlepage}

\begin{center}

\hfill February 2008\\

\vskip .5in

{\Large \bf
Non-Gaussianity, Spectral Index and Tensor Modes\\
  in Mixed Inflaton and Curvaton Models
}

\vskip .45in

{\large
Kazuhide Ichikawa$^1$,
Teruaki Suyama$^1$,
Tomo Takahashi$^2$ \\
and Masahide Yamaguchi$^3$
}

\vskip .45in

{\em
$^1$
Institute for Cosmic Ray Research, 
University of Tokyo, Kashiwa 277-8582, Japan\\
$^2$
Department of Physics, Saga University, Saga 840-8502, Japan \\
$^3$
Department of Physics and Mathematics, \\
Aoyama Gakuin University, Sagamihara 229-8558, Japan 
}

\end{center}

\vskip .4in

\begin{abstract}
 
  We study non-Gaussianity, the spectral index 
  of primordial scalar fluctuations and  tensor modes in models where
  fluctuations from the inflaton and the curvaton can both contribute to the
  present cosmic density fluctuations.  
  Even though simple single-field inflation models generate only 
  tiny non-Gaussianity,
  if we consider such a mixed scenario, large non-Gaussianity can be 
  produced.  Furthermore, we study the inflationary
  parameters such as the spectral index and the tensor-to-scalar ratio
  in this kind of models and discuss in what cases models
  predict the spectral index  and  tensor modes
  allowed by the current data while generating large non-Gaussianity, 
  which may have many implications for model-buildings of the inflationary
  universe.

\end{abstract}
\end{titlepage}

\renewcommand{\thepage}{\arabic{page}}
\setcounter{page}{1}
\renewcommand{\thefootnote}{\#\arabic{footnote}}

\section{Introduction}

It is widely believed that cosmic microwave background anisotropies and
large scale structure of the universe we observe today originate from
fluctuations generated during the time of inflation. It is often assumed
that the quantum fluctuation of the inflaton is responsible for that.
During inflation, the inflaton slowly rolls down its potential, which 
gives almost scale-invariant primordial fluctuations and some gravity waves can 
also be generated. Usually the primordial fluctuations are characterized by
the scalar spectral index and the
tensor-to-scalar ratio. Once the potential for the inflaton is
given, one can predict these quantities to compare with cosmological
observations. Current cosmological data are very precise to severely
constrain models of inflation and some models are considered to 
have been already excluded \cite{Spergel:2006hy,Tegmark:2006az}.

In fact, the considerations of the spectral index and tensor modes
may not be enough to probe primordial fluctuations.  
The non-Gaussianity of the fluctuations can also be measured
by cosmological observations and be used to test the scenario generating 
primordial fluctuations. In particular, 
there has recently been reported that
non-Gaussianity is detected in the cosmic microwave background
almost at 3$\sigma$ level \cite{Yadav:2007yy}.  
Since simple single-field inflation models predict very small 
non-Gaussianity,
if the evidence for large non-Gaussianity is mounting in the
future,
these models which at present are judged to be successful
as regards the amplitude and scale dependence for primordial
curvature fluctuations may suffer from a great difficulty.

However, notice that another source of fluctuations other than the inflaton 
can contribute to the cosmic density fluctuations.
Among such possibilities,   
the curvaton mechanism 
\cite{Enqvist:2001zp,Lyth:2001nq,Moroi:2001ct} and modulated reheating
scenarios \cite{Dvali:2003em,Kofman:2003nx}, for example, have been proposed and 
some of their observational consequences including 
issues of the non-Gaussianity have also been investigated. 
In particular, it is discussed that large non-Gaussianity can be 
produced in these scenarios.
It should be reminded that cosmic density fluctuations can originate from one of these
mechanisms including the inflaton or from some combination of these.
Thus,
even though almost perfect Gaussian fluctuations are generated in simple inflation 
models where they originate only from the inflaton,
large non-Gaussianity can be produced for total by adding the contribution
from another generation mechanism.
One of the purposes of this paper is to study the issue of non-Gaussianity 
in models of this kind, paying particular attention to  a mixed scenario where 
fluctuations of the inflaton and the curvaton 
can both contribute to primordial fluctuations.  
In investigating the non-Gaussianity,
we focus not only on the bispectrum but also on the trispectrum 
which is a potentially useful observable in the near future.

It should also be noticed that the predictions of the amplitude, the spectral index and 
tensor modes can be affected when another source of fluctuations is
introduced in the model.  Interestingly, some models of 
inflation which are disfavored by the data can be liberated 
by assuming the curvaton mechanism to work because of such modifications
\cite{Dimopoulos:2002kt,Endo:2003fr,Lazarides:2004we,Dimopoulos:2004yb,Rodriguez:2004yc,Langlois:2004nn,Moroi:2005kz,Moroi:2005np}. 
In particular, in Ref.~\cite{Moroi:2005kz}, it has been studied in what
cases models of inflation can be relaxed by adding fluctuations from the
curvaton in some detail assuming some concrete inflation models focusing
on the spectral index and the tensor-to-scalar ratio.  
In any case, since  we can expect much more precise measurements 
of these quantities in the future experiment such as Planck,  
we should  also investigate the predictions for the scale dependence of primordial
curvature fluctuations and tensor modes in the mixed scenario 
as well as that for non-Gaussianity. Another aim of this paper is to 
analyze this issue extending the work of Ref.~\cite{Moroi:2005kz}
to include the case where the curvaton is always subdominant component in the
history of the universe, which was not considered in
Ref.~\cite{Moroi:2005kz}. Then we study the effects of the curvaton on
inflationary parameters such as the spectral index and the tensor-to-scalar ratio 
and compare with recent observations of WMAP. 
Furthermore, we also discuss in what cases/models large non-Gaussianity 
can be generated satisfying the constraints on the scale dependence 
and tensor modes of primordial fluctuations in the scenario.
 
The structure of this paper is as follows. In the next section, we
discuss the formalism to study observational quantities in models with
mixed fluctuations from the inflaton and the curvaton.  
We derive the expressions for the scalar spectral index, 
its running, the tensor-to-scalar ratio and
the non-linearity parameters for such mixed models.  
Then, in Section
\ref{sec:obs}, we investigate those inflationary parameters 
assuming several concrete inflation models. 
First of all, we study 
the scalar spectral index and the tensor-to-scalar ratio
for each model and then discuss whether they are compatible with
current observations. For inflation models which are 
considered to have been already excluded by the data, 
we discuss in what case the curvaton can liberate the model.
Specifically, we discuss in what cases non-Gaussianity can be large.
The final section is devoted to conclusion and summary of this paper.

%%%%%%%%%%%%%%%%%%%%%%%%%%%%%%%
\section{Formalism}\label{sec:formalism}
%%%%%%%%%%%%%%%%%%%%%%%%%%%%%%%

%%%%%%%%%%%%%%%%%%%%%%%%%%%%%%%
\subsection{$\delta N$ formalism and some definitions}\label{sec:delta_N}
%%%%%%%%%%%%%%%%%%%%%%%%%%%%%%%

In the following, we adopt the $\delta N$ formalism
\cite{Starobinsky:1986fxa,Sasaki:1995aw,Sasaki:1998ug,Lyth:2004gb} to calculate the primordial power
spectrum and some non-linearity parameters. Thus we briefly review the
$\delta N$ formalism here.

In the $\delta N$ formalism, the primordial curvature perturbation
$\zeta$ on the uniform energy density hypersurface at the time $t=t_f$
is equal to the
perturbation in the local expansion defined with respect to an initial
spatially flat hypersurface.  On sufficiently large scales where the
spatial gradient can be neglected, the local expansion is well
approximated by the expansion of the unperturbed universe,
\begin{equation}
%\label{ }
N (t_\ast, t_f, x) = \int_{t_\ast}^{t_f} H(x,t) dt. 
\end{equation}
where $H$ is the local Hubble expansion and $t_\ast$ is some time during
inflation. Then the primordial curvature perturbation can be expressed
as
\begin{eqnarray}
\zeta (t_f,{\vec x}) =N(t_f, t_*, {\vec x}) - \bar{N}, 
\end{eqnarray}
where $\bar{N}$ is the expansion in the background spacetime:
\begin{equation}
%\label{ }
\bar{N} = \int_{t_\ast}^{t_f} \bar{H}(t) dt. 
\end{equation}
If we take $t_\ast$ as a time when the cosmological scale crossed the
horizon scale during inflation, then $N(t_f,t_\ast,{\vec x})$ becomes a
function of the scalar field at the time of horizon crossing.  Hence
$\zeta$ can be written as,
\begin{eqnarray}
\zeta (t_f) = N_a \delta \phi^a_\ast
+\frac{1}{2}N_{ab} \delta \phi_*^a \delta \phi_*^b
+\frac{1}{6}N_{abc} \delta \phi_*^a \delta \phi_*^b \delta \phi_*^c 
+ \cdots,
\label{eq:zeta_expansion}
\end{eqnarray}
where the summation is implied over repeated indices which label the
scalar field. Here $\delta \phi_*^a$ is the perturbation of the scalar
field $\phi^a$ on the flat slicing at the time of horizon crossing. In
the following, when the asterisk $\ast$ is shown in the subscript, it
indicates that the quantities are evaluated at the time of horizon
crossing. $N_a$, $N_{ab}$, and $N_{abc}$ are given by
\begin{equation}
%\label{ }
N_a \equiv \frac{\partial N}{\partial \phi^a}, ~~~
N_{ab} \equiv \frac{\partial^2 N}{\partial \phi^a \partial \phi^b}, ~~~ 
N_{abc} \equiv \frac{\partial^3 N}{\partial \phi^a \partial \phi^b \partial \phi^c}.
\end{equation}
For the purpose of this paper, we include the terms up to cubic order in
the perturbations of the scalar field.  If we choose $t_f$ well after the
reheating, then $\zeta (t_f)$ gives the primordial adiabatic
perturbations.

Now we write down the expression for the primordial power spectrum
which is defined as
\begin{equation}
\label{eq:power}
\langle \zeta_{\vec k_1} \zeta_{\vec k_2} \rangle 
= 
{(2\pi)}^3 P_\zeta (k_1) \delta ({\vec k_1}+{\vec k_2}).
\end{equation}
By using Eq.~\eqref{eq:zeta_expansion}, we can express $P_\zeta$ with
the fluctuations of scalar fields as
\begin{equation}
P_\zeta (k)
=
N_a N_b P^{ab} (k), 
\label{eq:P_zeta} 
\end{equation}
where $P^{ab} (k)$ is the power spectrum of scalar fields:
\begin{equation}
\label{eq:P_phi}
\langle \delta \phi^a_{\vec k_1} \delta \phi^b_{\vec k_2} \rangle
=
{(2\pi)}^3 P^{ab} (k_1) \delta ({\vec k_1}+{\vec k_2}).
\end{equation}
In this paper, we consider two scalar fields, the inflaton and the curvaton
which are denoted as $\phi$ and $\sigma$ and their fluctuations are
indicated as $\delta \phi$ and $ \delta \sigma$ respectively. Thus the
indices for scalar fields are understood to represent either of these
fields in the following.  The fluctuations at the time of horizon
crossing $\delta \phi_\ast$ and $\delta \sigma_\ast$ are assumed to be
uncorrelated random fields with the same amplitude. 
Hence we have
\begin{eqnarray}
P^{ab} (k)= P(k) \delta^{ab}=\frac{2\pi^2}{k^3} \left( \frac{H_\ast}{2 \pi} \right)^2 \delta^{ab}. \label{uncorrelated}
\end{eqnarray}
Then at leading order in fluctuations of the scalar fields, 
the power spectrum can be written as
\begin{equation}
\label{eq:P_ab}
P_\zeta (k) = \frac{2\pi^2}{k^3}{\cal P}_\zeta (k)=N_a N^a \frac{2\pi^2}{k^3} \left( \frac{H_\ast}{2 \pi} \right)^2.
\end{equation}
The expression for $N_a$ will be discussed in the next subsection.

Now we discuss quantities which represent non-Gaussianity, the
bispectrum and trispectrum. 
The definition of the bispectrum $B_\zeta$ is given by
\begin{eqnarray}
\langle \zeta_{\vec k_1} \zeta_{\vec k_2} \zeta_{\vec k_3} \rangle
&=&
{(2\pi)}^3 B_\zeta (k_1,k_2,k_3) \delta ({\vec k_1}+{\vec k_2}+{\vec k_3}).
\label{eq:bi} 
\end{eqnarray}
As is the case with the power spectrum, 
the leading order bispectrum can be written  as
\begin{eqnarray}
B_\zeta (k_1,k_2,k_3)
&=&
N_a N_b N_c B^{abc} (k_1,k_2,k_3) \notag \\
&& 
+N_a N_{bc} N_d  
\left( 
P^{ac}(k_1) P^{bd} (k_2)
+ P^{ac}(k_2) P^{bd} (k_3) 
+ P^{ac}(k_3) P^{bd} (k_1)
 \right),
\label{eq:B_zeta} 
\end{eqnarray}
where $B^{abc}$ is the bispectrum of the scalar fields 
\begin{equation}
\label{eq:bi_phi}
\langle 
\delta \phi^a_{\vec k_1} \delta \phi^b_{\vec k_2} \delta \phi^c_{\vec k_3} 
\rangle
={(2\pi)}^3 B^{abc} (k_1,k_2,k_3) \delta ({\vec k_1}+{\vec k_2}+{\vec k_3}).
\end{equation}

The trispectrum $T_\zeta$ is defined as
\begin{eqnarray}
\langle 
\zeta_{\vec k_1} \zeta_{\vec k_2} \zeta_{\vec k_3} \zeta_{\vec k_4} 
\rangle
&=&
{(2\pi)}^3 T_\zeta (k_1,k_2,k_3,k_4) \delta ({\vec k_1}+{\vec k_2}+{\vec k_3}+{\vec k_4}),
\label{eq:tri}
\end{eqnarray}
and can be given with the power spectrum $P^{ab}$, the bispectrum
$B^{abc}$ and the trispectrum $T^{abcd}$ for the scalar fields,
\begin{eqnarray}
T_\zeta (k_1,k_2,k_3,k_4)
&=&
N_a N_b N_c N_d T^{abcd}(k_1,k_2,k_3,k_4) \notag \\
&& 
+ N_{ab} N_c N_d N_e \left( P^{ac}(k_1) B^{bde} (k_{12},k_3,k_4)+11~{\rm perms.} \right) \notag \\
&&
+N_{ab}N_{cd} N_e N_f 
\left( P^{bd}(k_{13})P^{ae}(k_3)P^{cf}(k_4)+11~{\rm perms.} \right) \notag \\
&&
+N_{abc}N_d N_e N_f \left( P^{ad}(k_2)P^{be}(k_3) P^{cf}(k_4)+3~{\rm perms.} \right), \label{eq:T_zeta}
\end{eqnarray}
where $k_{ij} = | k_i + k_j|$ and $T^{abcd}$ is given by
\begin{equation}
\label{eq:tri_phi}
\langle 
\delta \phi^a_{\vec k_1} \delta \phi^b_{\vec k_2} 
\delta \phi^c_{\vec k_3} \delta \phi^d_{\vec k_4} 
\rangle 
=
{(2\pi)}^3 T^{abcd} (k_1,k_2,k_3,k_4) 
\delta ({\vec k_1}+{\vec k_2}+{\vec k_3}+{\vec k_4}).
\end{equation}
Expressions for $B_{abc}$ and $T_{abcd}$ were provided in
\cite{Seery:2005gb} and \cite{Seery:2006vu} for general multiple-fields
slow-roll inflation (see also \cite{Arroja:2008ga}).
In the following, we will
neglect $B_{abc}$ and $T_{abcd}$ because the inclusion of these terms only
gives corrections of slow-roll order to the primordial bispectrum and
trispectrum, which are far below the observational sensitivity expected
for the Planck satellite. Then $B_\zeta$ and $T_\zeta$ can be written
as
\begin{eqnarray}
B_\zeta (k_1,k_2,k_3)
&=&
\frac{6}{5} f_{\rm NL} 
\left( 
P_\zeta (k_1) P_\zeta (k_2) 
+ P_\zeta (k_2) P_\zeta (k_3) 
+ P_\zeta (k_3) P_\zeta (k_1)
\right), \\
\label{eq:def_f_NL}
T_\zeta (k_1,k_2,k_3,k_4)
&=&
\tau_{\rm NL} \left( 
P_\zeta(k_{13}) P_\zeta (k_3) P_\zeta (k_4)+11~{\rm perms.} 
\right) \nonumber \\
&&
+ \frac{54}{25} g_{\rm NL} \left( P_\zeta (k_2) P_\zeta (k_3) P_\zeta (k_4)
+3~{\rm perms.} \right),
\label{eq:def_tau_g_NL} 
\end{eqnarray}
where $f_{NL},~\tau_{NL}$ and $g_{NL}$ are constant parameters 
which are given in terms of the derivatives of the number of $e$-folding
with respect to the scalar fields as \cite{Lyth:2005fi,Alabidi:2005qi,Byrnes:2006vq},
\begin{eqnarray}
\frac{6}{5}f_{\rm NL}
&=&
 \frac{N_a N_b N^{ab}}{ {(N_c N^c)}^2 }, 
\label{eq:f_NL_N} \\
\tau_{\rm NL}
&=&
\frac{N_{ab} N^{ac} N^b N_c}{ {(N_d N^d)}^3 }, 
\label{eq:tau_NL_N} \\
\frac{54}{25} g_{\rm NL}
&=&
\frac{N_{abc} N^a N^b N^c}{ {(N_d N^d)}^3 }. 
\label{eq:g_NL_N}
\end{eqnarray}
These expressions are valid for the case where $B_{abc}$ and $T_{abcd}$ 
are negligibly small. Furthermore, an interesting inequality can be
obtained with the help of the Cauchy-Schwartz inequality \cite{Suyama:2007bg}:
\begin{equation}
\label{eq:inequality}
\tau_{\rm NL} \ge \frac{36}{25} f_{\rm NL}^2.
\end{equation}
This inequality holds for the scenarios such as the curvaton
and the modulated reheating scenarios where the
leading non-Gaussianity comes from super-horizon evolution.

To calculate the primordial power spectrum and the non-linearity
parameters, we need to know the number of $e$-folding as a function of
the field value of the scalar fields, the inflaton and the curvaton,
from the time when the cosmological scales crossed the horizon to
the time when the universe becomes radiation-dominated connected to big bang
nucleosynthesis (BBN) \footnote{
  Since we consider the scenario with the curvaton, the universe could
  have experienced the dominance of radiation twice after inflation
  ends: radiations from the decay of inflaton and curvaton.
  Radiation-dominated epoch here is meant to be the radiation-dominated
  phase after the curvaton decay.
}. In the next subsection, we discuss this issue in detail.

%%%%%%%%%%%%%%%%%%%%%%%%%%%%%%
\subsection{Background dynamics and the number of $e$-folding}
%%%%%%%%%%%%%%%%%%%%%%%%%%%%%%

Here we give an explicit expression for the number of $e$-folding in
models with the inflaton and the curvaton. For the potential of the
curvaton, we take a quadratic potential, 
\begin{equation}
\label{eq:V_sigma}
U(\sigma) = \frac{1}{2} m_\sigma^2 \sigma^2,
\end{equation}
where $m_\sigma$ is the curvaton mass.  We assume that the energy
density of the curvaton is subdominant during inflation and at least
until the time when the inflaton decays into radiation. In addition, the
the curvaton mass is assumed as $m_\sigma \ll H_{\rm inf}$ with
$H_{\rm inf}$ being the Hubble parameter during inflation. In this case,
the curvaton field almost stays at the initial value during inflation.
After inflation ends, the inflaton oscillates around the minimum of its
potential for some time, then decays into radiation. The curvaton also
begins to oscillate {at around a time when $H=m_\sigma$ and decays into
radiation.  In the following, we assume that the curvaton
decays long after the onset of the curvaton oscillations, which is
equivalent to assume that $ \Gamma_\sigma/ m_\sigma \ll 1$ with $\Gamma_\sigma$
being the decay rate of the curvaton.

Here we give a rough sketch of the thermal history of the universe in
this scenario. In fact, depending on the initial amplitude of the curvaton
field, the thermal history after the inflaton decay can become
different.  When the initial value of the curvaton $\sigma_{\rm in}$ is
small enough, typically as $\sigma_{\rm in} \ll M_{\rm pl}$, the
curvaton begins to oscillate around the minimum of the potential during
radiation-dominated epoch due to the inflaton decay. With the potential
Eq.~\eqref{eq:V_sigma}, the energy density of the curvaton decreases as
$\rho_\sigma \propto a^{-3}$ when $\sigma$ oscillates. On the other
hand, the energy density of radiation decreases as $\rho_{\rm rad}
\propto a^{-4}$, thus the curvaton can become dominant if the curvaton
oscillates around the potential minimum long enough before decaying into
radiation.  If the decay rate of the curvaton is not so small, it can
decay before it becomes a dominant component.  In any case, when the
rate of the Hubble expansion $H$ becomes comparable to the decay
rate of the curvaton, the curvaton decays into radiation.  After the
curvaton decays into radiation, the universe is radiation-dominated and
is led to BBN epoch.

When the initial value for the curvaton field is large enough, typically
as $\sigma_{\rm in} \gg M_{\rm pl}$, 
the curvaton almost stays at the initial value and drives the second 
inflation even after the onset of the radiation-dominated epoch due to
the inflaton decay.
The thermal history in this case is as follows: after the inflation
driven by the inflaton ends, the inflaton decays into radiation, then
the universe becomes radiation dominated. When the initial amplitude of
the curvaton is large, $\sigma$ stays at the initial value until long
after the inflaton decays into radiation. Since the energy density of
such a scalar field is constant, at some epoch, the energy density of
the curvaton dominates to give the second inflationary epoch. This
situation is similar to that of a double inflation model
\cite{Polarski:1992dq}. After the curvaton drives the second inflation,
it begins to oscillate around the minimum of the potential when the
Hubble parameter decreases as $H \sim m_\sigma$.  Then the curvaton
decays into radiation when $\Gamma_\sigma \sim H$.

Since we would like to calculate the primordial power spectrum and some
non-linearity parameters during the radiation-dominated epoch after the
curvaton decay, we need to know the $e$-folding number from the time
when the cosmological scale exits the horizon during inflation to the
time after the decay of the curvaton.  
We denote the number of $e$-folding during these epochs as $N_{\rm tot}$.  
For later convenience, 
we divide $N_{\rm tot}$ into several parts as
\begin{equation}
\label{eq:N_tot}
N_{\rm tot} = N_{\rm inf}  + N_{\rm d}  +N_{\rm R} 
\end{equation} 
where $N_{\rm inf}$ represents the number of $e$-folding during
inflation driven by the inflaton. $N_{\rm d}$ is for the epoch from the
end of inflation to some time after the decay of the inflaton.  
$N_{\rm R}$ refers to the $e$-folding number from that time to the time well
after the curvaton decay.  $N_{\rm R}$ is relevant to the curvaton
dynamics.

Now we make detailed discussion on $N_{\rm inf}, N_{\rm d}$ and 
$N_{\rm R}$ in order to calculate the primordial power spectrum and the
non-linearity parameters in the $\delta N$ formalism.  We do not discuss
$N_{\rm d}$ here since it is irrelevant to the primordial
fluctuation\footnote{
In the case that the curvaton energy density
becomes dominant during the regime of the inflaton oscillations, 
$N_{\rm d}$ depends on $\sigma_\ast$ so that it is also relevant to the primordial
fluctuations. However, as stated in the paragraph below Eq.~(\ref{eq:V_sigma}), 
we do not consider such a case in
this paper. 
}. However, when we
relate the cosmological scales observed today to the time of horizon
exit during inflation, we also need $N_d$, which we will discuss in the
subsection~\ref{sec:scale}.

First we start with $N_{\rm inf}$,
the number of $e$-folding during inflation.
It is determined by the dynamics of the inflaton whose
equation of motion is
\begin{equation}
%\label{ }
\ddot{\phi} + 3 H \dot{\phi} + V_\phi = 0,
\end{equation} 
where a dot represents the derivative with respect to the cosmic time 
and $V_\phi =d V / d \phi$. 
During the inflation, 
the so-called slow-roll (SR) approximation is valid.
Hence $N_{\rm inf}$ can be written as, by using
$\dot{\phi} \simeq - V_\phi /(3H)$,
\begin{equation}
\label{eq:N_inf}
N_{\rm inf} = \int_{t_\ast}^{t_{\rm e}} H dt  
\simeq  
- \frac{1}{M_{\rm pl}^2} \int_{\phi_\ast}^{\phi_{\rm e}}  \frac{V}{V_\phi} d \phi
\end{equation}
where $t_\ast$ and $t_{\rm e}$ respectively represent the times 
when a scale exits the horizon and when the inflation ends. 
Similarly, 
$\phi_\ast$ and $\phi_{\rm e}$ indicate the scalar field values at corresponding
epochs. To give a more concrete expression for $N_{\rm inf}$, 
we need to specify the potential for the inflaton.

Next we consider $N_{\rm R}$.  
After the inflaton decays into the
radiation, the universe is composed of radiation and the curvaton.
Hence the background equations are given by
\begin{eqnarray}
&&{\dot \rho}_r+4H\rho_r 
= \Gamma_\sigma \rho_\sigma, 
\label{eq:rad} \\
&&{\ddot \sigma} +(3H+\Gamma_\sigma) {\dot \sigma}+m_\sigma^2 \sigma = 0, 
\label{eq:curvaton_eom} \\
&&H^2 = \frac{1}{3M_{\rm pl}^2} (\rho_r+\rho_\sigma), 
\label{eq:friedmann}
\end{eqnarray}
where $\rho_r$ and $\rho_\sigma = (1/2) \dot{\sigma}^2 + U(\sigma)$
are energy densities of radiation and the curvaton, respectively.
To obtain $N_{\rm R}$, 
we solve the above equations from the time $t = t_0$ which
corresponds to some time after the reheating epoch due to the 
decay of the inflaton with the initial conditions
\begin{eqnarray}
\rho_r (t_0)=\rho_{r0},~~~\sigma (t_0)=\sigma_\ast,
~~~{\dot \sigma}(t_0) \simeq -\frac{m_\sigma^2 \sigma_\ast}{3H(t_0)},
\end{eqnarray}
to the time $t_f$ when the Hubble parameter becomes much smaller than
$\Gamma_\sigma$, i.e.,$H(t_f) \ll \Gamma_\sigma$.  Here we denote the
initial amplitude for the curvaton field with an asterisk as
$\sigma_\ast$. Remind that we are using an asterisk to represent that a
quantity is evaluated at the time of horizon exit.  Since $\sigma$
almost stays at the initial position during inflation, the initial value
for $\sigma$ here is the same as that at the horizon exit during
inflation.  By integrating Eq.~\eqref{eq:rad}, we can write $N_{\rm R}$
formally as
\begin{equation}
N_{\rm R} 
=\frac{1}{4} \log \frac{\rho_{r0}}{\rho_f}+\frac{1}{4} 
\log \left( 1+F(\sigma_\ast,m_\sigma,\Gamma_\sigma) \right),
\label{eq:N_cur}
\end{equation}
where $F(\sigma_\ast, m_\sigma, \Gamma_\sigma)$ is defined as
\begin{equation}
F(\sigma_\ast,m_\sigma,\Gamma_\sigma) 
\equiv 
\int_0^{\infty}dN~e^{4N}\frac{\Gamma_\sigma}{H(N)} \frac{\rho_\sigma (N)}{\rho_{r0}},
\label{eq:def_Q}
\end{equation}
and $\rho_f =\rho_r(t_f)$.  
In general, we need a numerical calculation
to evaluate $F(\sigma_\ast,m_\sigma,\Gamma_\sigma)$. 
However, for some cases, 
we can integrate analytically the right hand side (RHS)
of Eq.~(\ref{eq:def_Q}), which will be explicitly
given in the next subsection.  
It should be noted here that,
when the curvaton does not drive the secondary inflation and begins to
oscillate during the radiation-dominated epoch caused by the inflaton decay,
the value of $F$ depends only on the combination 
\begin{equation}
\label{eq:def_p}
p \equiv \frac{\sigma_\ast^2}{M_{\rm pl}^2\sqrt{\Gamma_\sigma/m_\sigma}}.
\end{equation}

When the curvaton is oscillating, it can be regarded as a matter fluid
as long as the background evolution is concerned. Then the universe 
can be regarded as a two component system which consists of matter and
radiation. In this case, it has been shown that the solution for the
system of Eqs.~\eqref{eq:rad}, \eqref{eq:curvaton_eom} and
\eqref{eq:friedmann} depends only on the single parameter $p$ defined
above \cite{Malik:2002jb}.  Thus $F$ can be determined once $p$ is fixed
for such a case.

When the second inflation driven by the curvaton occurs, the curvaton
cannot be regarded as matter. However, even in this case, the value of
$F$ can be determined once $\sigma_\ast$ and the combination
$\Gamma_\sigma / m_\sigma$ are given.  
During the second inflationary phase, 
the slow-roll approximation is valid for the curvaton field as in
the case during inflation.
In this case,
$N_{\rm R}$ can be written as
\begin{equation}
\label{eq:N_cur2}
N_{\rm R} \simeq -\frac{1}{M_{\rm pl}^2}\int_{\sigma_\ast}^{\sigma_{\rm end}}  \frac{U}{U_\sigma} d \sigma 
+ C,
\end{equation}
where $\sigma_{\rm end}$ is the value of $\sigma$
at the end of the second inflation. 
$C$ represents the $e$-folding number from the end of the
second inflation to the time well after the curvaton decay
and does not depend on $\sigma_*$.
Thus, even in this case, once the
combination $\Gamma_\sigma/m_\sigma$ is given along with
$\sigma_\ast$, the number of $e$-folding $N_{\rm R}$ is determined.
Hence,
we define the combination of the variables as
\begin{equation}
\label{eq:s}
s \equiv \frac{\Gamma_\sigma}{m_\sigma},
\end{equation}
which will be used frequently instead of giving $\Gamma_\sigma$ and $m_\sigma$
separately. 
As stated before,
we assume $s \ll 1$ in this paper.

Having described the necessary formulae in the $\delta N$ formalism, we
discuss primordial curvature fluctuations, tensor modes and
non-Gaussianity in the following subsections.

%%%%%%%%%%%%%%%%%%%%%%%%%%%%%%
\subsection{Primordial power spectrum}
%%%%%%%%%%%%%%%%%%%%%%%%%%%%%%

Now we discuss primordial power spectrum in models with mixed
fluctuations from the inflaton and the curvaton. By using the $\delta N$
formalism, the primordial curvature perturbation $\zeta$ can be
calculated as
\begin{equation}
\label{eq:zeta1}
\zeta \simeq \frac{1}{M_{\rm pl}^2}\frac{V}{V_\phi} \delta \phi_\ast 
+   \frac{\partial Q}{\partial \sigma} \delta \sigma_\ast,
\end{equation}
where $Q$ represents the $\sigma_\ast$ dependent part of $N_{\rm R}$
which is defined as
\begin{equation}
\label{eq:F}
Q \equiv \frac{1}{4} \log (1+F).
\end{equation}
In actual calculations, we numerically obtain the function $F$ in all
cases.  The primordial power spectrum for a model with mixed inflaton
and curvaton fluctuations is given as
\begin{equation}
\label{eq:P_zeta2}
{\cal P}_\zeta 
= 
\left(  \frac{1}{M_{\rm pl}^4} \frac{V^2}{V_\phi^2}  + Q_\sigma^2 \right) \left( \frac{H}{2\pi} \right)^2,
\end{equation}
where $Q_\sigma \equiv \partial Q / \partial \sigma$ and we used
Eq.~\eqref{uncorrelated}. The spectral index and its running are given
by
\begin{eqnarray}
\label{eq:n_s}
n_s - 1 
& \equiv & 
 \left. \frac{d \log {\cal P}_\zeta }{ d \log k} \right|_{k = aH }  \notag \\
 &=&
 -2 \epsilon - \frac{ 4 \epsilon - 2 \eta }{1 + 2 \epsilon M_{\rm pl}^2 Q_\sigma^2}, \\
\label{eq:n_run}
 n_{\rm run} 
 &\equiv &
\frac{d n_s}{d \log k} \notag \\
&=&
-4 \epsilon (2 \epsilon - \eta ) - \frac{2}{ (1 + 2 \epsilon M_{\rm pl}^2 Q_\sigma^2)^2}
\left[ 
8 \epsilon^2 - 6 \epsilon \eta + \xi^2 + 
2 \epsilon ( 2 \epsilon \eta - 2 \eta^2  + \xi^2) M_{\rm pl}^2 Q_\sigma^2 
\right], \notag \\
\end{eqnarray}
where $\epsilon, \eta$ and $\xi^2$ are the slow-roll parameters which
are defined as
\begin{equation}
\label{eq:slow_roll}
  \epsilon \equiv \frac{1}{2} M_{\rm pl}^2 
  \left( \frac{V_\phi}{V} \right)^2,
  ~~~
  \eta \equiv M_{\rm pl}^2 \frac{V_{\phi\phi}}{V},
  ~~~
   \xi^2 \equiv M_{\rm pl}^4
  \frac{V_\phi {V_{\phi\phi\phi} }}{V^2}.
\end{equation}
Here the third parameter $\xi^2$ should be considered to be the second
order in the slow-roll.

During inflation, tensor modes are also generated as
\begin{equation}
\label{eq:tensor}
{\cal P}_T = \frac{8}{M_{\rm pl}^2} \left( \frac{H}{2\pi} \right)^2. 
\end{equation}
When one extracts the information on tensor modes, the
tensor-to-scalar ratio is usually used, which is defined and given in
the case considered here as
\begin{equation}
\label{eq:r}
r \equiv \frac{{\cal P}_T}{{\cal P}_\zeta} 
= 
\frac{16 \epsilon}{1 + 2 \epsilon M_{\rm pl}^2 Q_\sigma^2}.
\end{equation}

In the pure curvaton limit, 
$2 \epsilon M_{\rm pl}^2 Q_\sigma^2 \gg 1$,
where the primordial fluctuations
are completely dominated by the curvaton fluctuations,
$n_S$, $n_{\rm run}$ and $r$ reduce to
\begin{eqnarray}
&&n_s-1=-2\epsilon, \\
&&n_{\rm run}=-4 \epsilon (2\epsilon-\eta), \\
&&r=\frac{8}{M_{\rm pl}^2 Q_\sigma^2}.
\end{eqnarray}
We find that the spectrum of the scalar fluctuations becomes red-tilted, 
the running of the spectral index becomes independent of
$\xi^2$, and the tensor-to-scalar ratio becomes negligible.

%%%%%%%%%%%%%%%%%%%
\subsection{Non-linearity parameters}
%%%%%%%%%%%%%%%%%%%

In the subsection \ref{sec:delta_N}, we gave the definitions of
non-linearity parameters of $f_{\rm NL}, \tau_{\rm NL}$ and $g_{\rm NL}$
in the $\delta N$ formalism. Here we give concrete expressions for them
in models with mixed fluctuations from the inflaton and the curvaton.

To obtain the bispectrum and the trispectrum,
we need to expand $\delta N$ up to cubic order in $\delta \phi$
and $\delta \sigma$,
which can be written as
\begin{eqnarray}
\label{eq:zeta2}
\zeta 
\simeq 
\frac{V}{V_\phi}  \delta \phi_\ast
+ \frac{1}{2} \left( 1 - \frac{VV_{\phi\phi}}{V_\phi^2} \right) \delta \phi_\ast^2
-\frac{V_{\phi\phi}}{6V_\phi} \left( 1+\frac{V V_{\phi\phi\phi}}{V_\phi V_{\phi\phi}}-\frac{2VV_{\phi\phi}}{V_\phi2} \right) \delta \phi_\ast^3 \nonumber \\
+ Q_\sigma \delta \sigma_\ast
+ \frac{1}{2} Q_{\sigma\sigma} \delta \sigma_\ast^2+\frac{1}{6} Q_{\sigma\sigma\sigma} \delta \sigma_\ast^3.
\end{eqnarray}
Then,
using Eqs.~(\ref{eq:f_NL_N}), (\ref{eq:tau_NL_N}) and
(\ref{eq:g_NL_N}),
the non-linearity parameters are given by}
\begin{eqnarray}
\label{eq:f_NL}
&&
\frac{6}{5} f_{\rm NL} 
=
\frac{1}{ (1+ 2 \epsilon M_{\rm pl}^2 Q_\sigma^2)^2} 
\left[
2 \epsilon -\eta + 4\epsilon^2 M_{\rm pl}^4 Q_\sigma^2 Q_{\sigma\sigma} 
\right], \\ 
\label{eq:tau_NL}
&&
\tau_{\rm NL}  = 
\frac{1}{ (1+ 2 \epsilon M_{\rm pl}^2 Q_\sigma^2)^3} 
\left[ ( 2 \epsilon -\eta)^2  + 8\epsilon^3 M_{\rm pl}^6 Q_\sigma^2 Q_{\sigma\sigma}^2 \right], \\
\label{eq:g_NL}
&&
\frac{54}{25}g_{\rm NL}  = 
\frac{1}{ (1+ 2 \epsilon M_{\rm pl}^2 Q_\sigma^2)^3} 
\left[  
- 2 \epsilon \eta  - \xi^2 + 2\eta^2  
+ 8 \epsilon^3 M_{\rm pl}^6 Q_\sigma^3 Q_{\sigma\sigma\sigma} 
\right].
\end{eqnarray}

In the pure curvaton limit,
$2\epsilon M_{\rm pl}^2 Q_\sigma^2 \gg 1$,
these expressions reduce to
\begin{eqnarray}
&&\frac{6}{5}f_{\rm NL}=\frac{Q_{\sigma \sigma}}{Q_\sigma^2}, \\
&&\tau_{\rm NL}=\frac{Q_{\sigma \sigma}^2}{Q_\sigma^4}=\frac{36}{25} f_{\rm NL}^2, \\
&&\frac{54}{25}g_{\rm NL}=\frac{Q_{\sigma \sigma \sigma}}{Q_\sigma^3}.
\end{eqnarray}

\begin{figure}[htb]
\begin{center}
\scalebox{0.6}{
\includegraphics{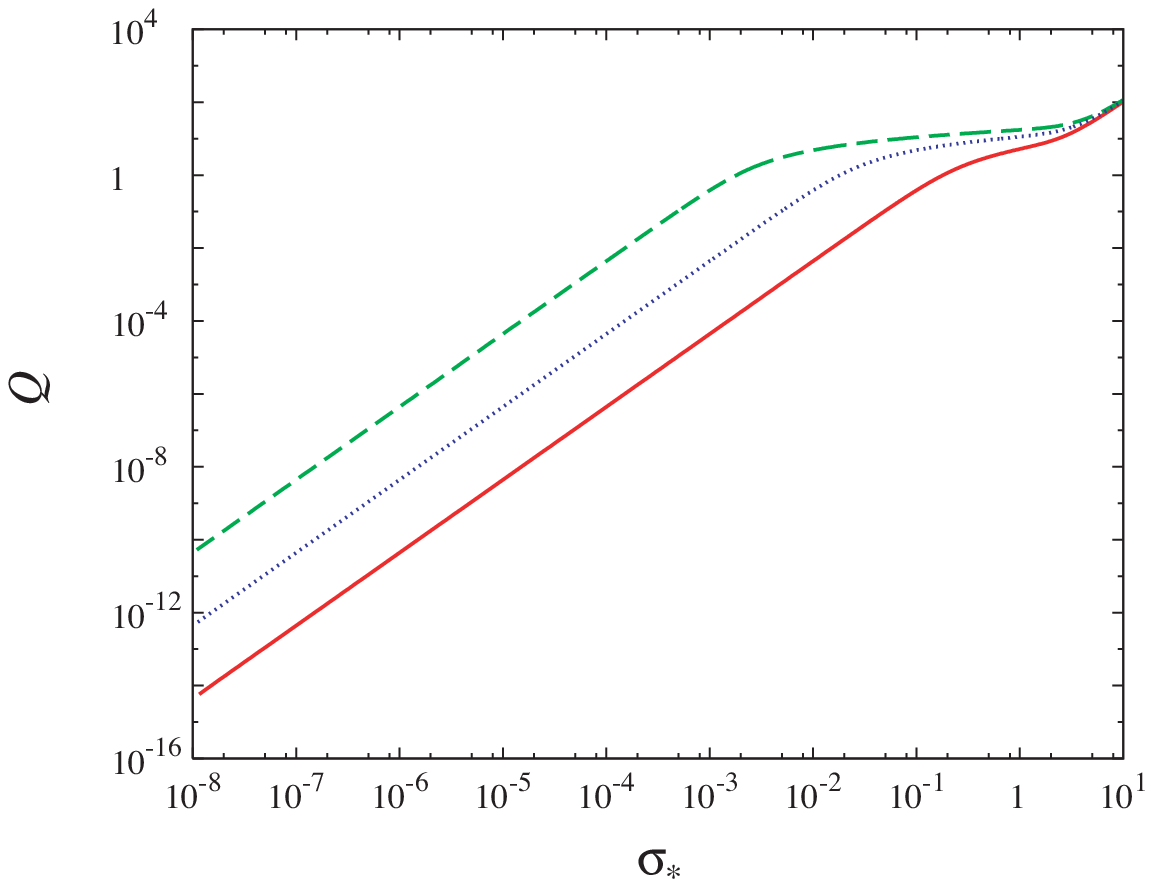} 
\includegraphics{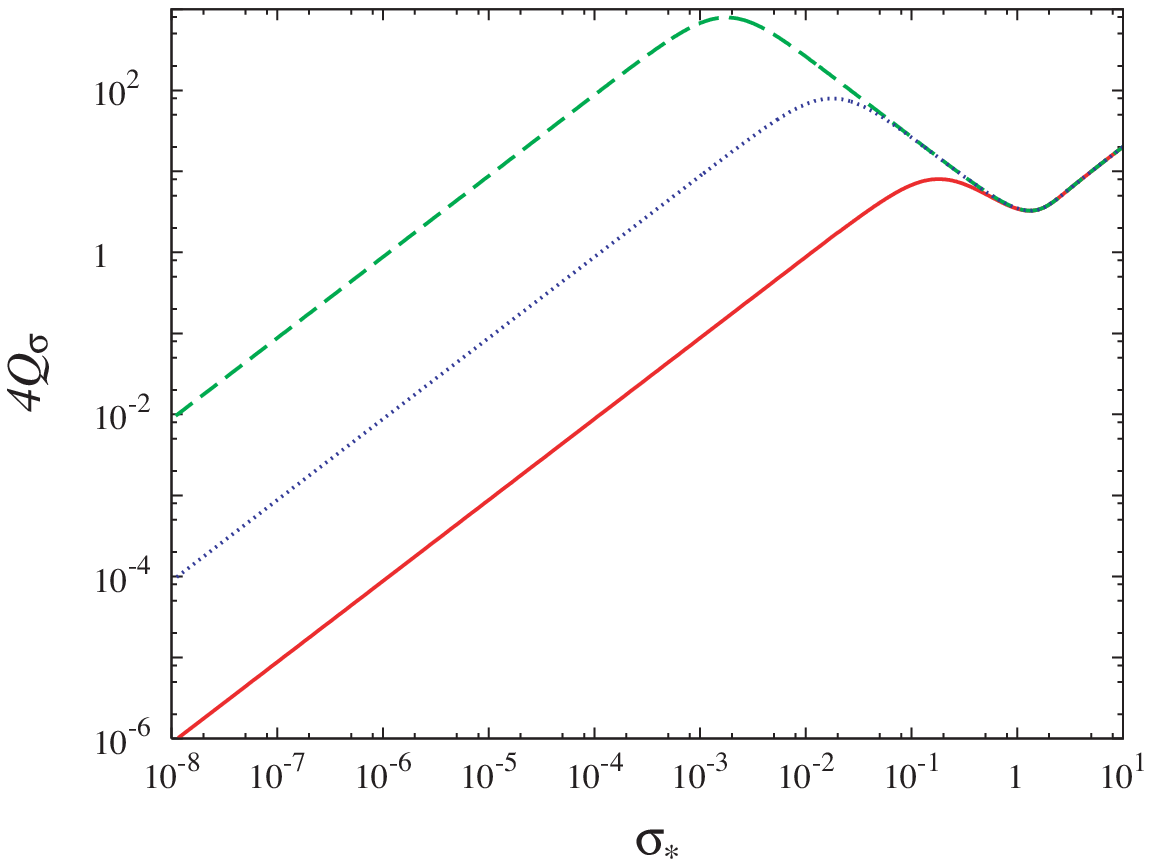}
}\\
\scalebox{0.6}{
\includegraphics{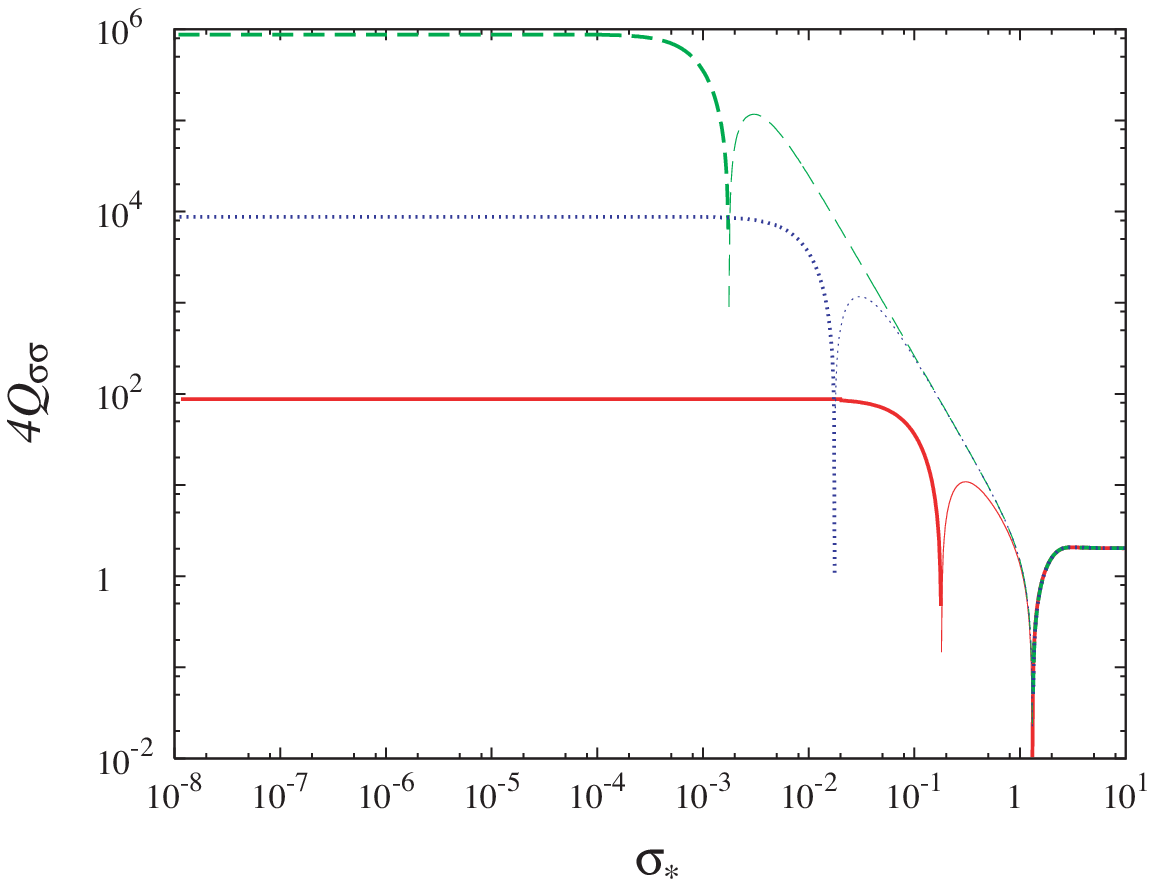}
\includegraphics{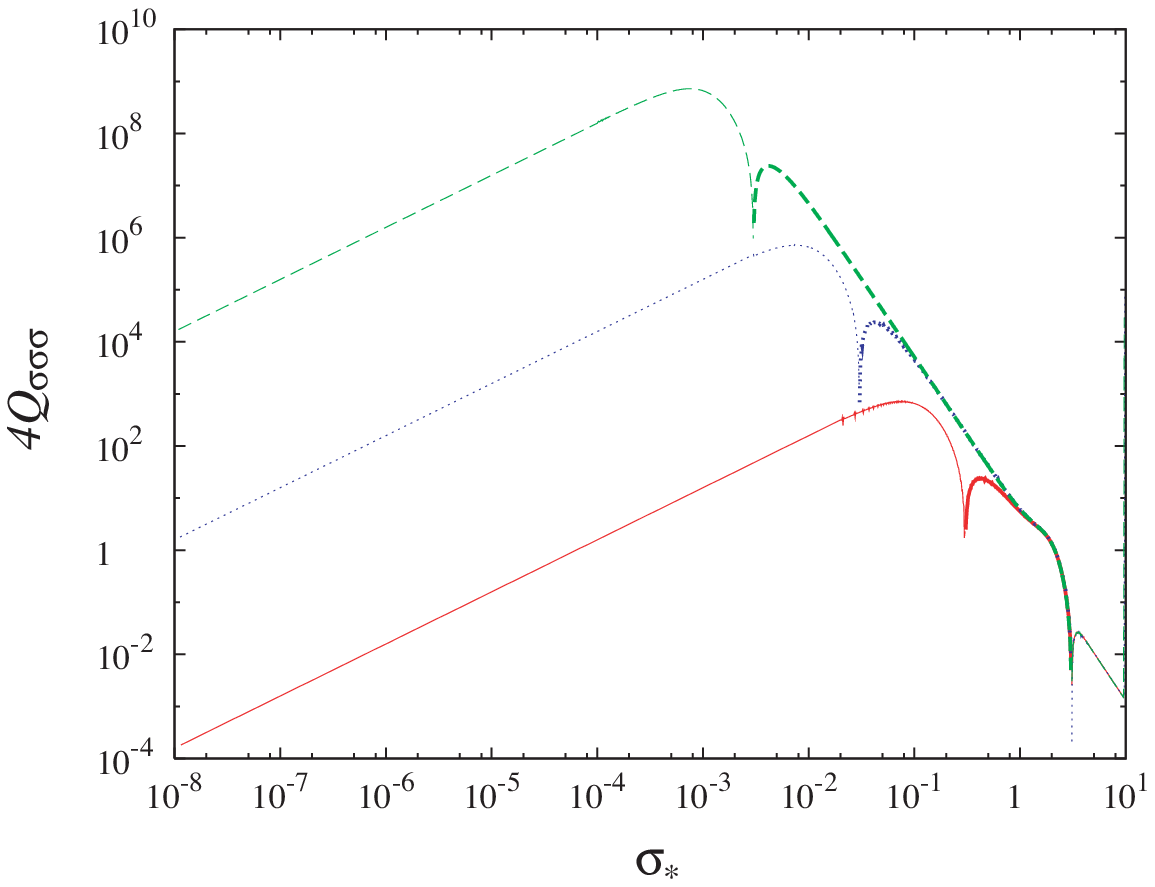}
}
\caption{Plots of $Q$ (top left), $Q_\sigma,$ (top right),
$Q_{\sigma \sigma}$ (bottom right) and $Q_{\sigma \sigma \sigma}$ (bottom left) 
as functions of $\sigma_\ast$ 
for the cases with $s=10^{-4}$ (red solid line), $s=10^{-8}$ (blue dotted line)
and $s=10^{-12}$ (green dashed line). Notice that $Q_{\sigma \sigma}$
and $Q_{\sigma \sigma \sigma}$ can be negative.  
When the functions take a negative value, 
the absolute value is drawn with a thin line.
$\sigma_*$ is shown in units of $M_{\rm pl}$.}
\label{fig:Q}
\end{center}
\end{figure}

Once we pick up a model of inflation and calculate
the values of $Q_\sigma, Q_{\sigma\sigma}$ and 
$Q_{\sigma\sigma\sigma}$, we can give  concrete values for these 
non-linearity parameters along with $n_s$ and $r$. 
Although an analytic expression can be obtained for 
some limiting values of $p$ and $\sigma_*$, 
which is going to be discussed in the next subsection, 
a numerical calculation is needed 
to evaluate $Q_\sigma, Q_{\sigma\sigma}$ and 
$Q_{\sigma\sigma\sigma}$ in general.
In Fig.~\ref{fig:Q}, we show them as functions of $\sigma_\ast$ 
for several values of $s$. From the figure, we can obtain some idea 
of when the contribution from the curvaton significantly affects 
the primordial power spectrum and non-Gaussianity.

%%%%%%%%%%%%%%%%%%%%%%%%%%%%%%%%%%%%%%
\subsection{Analytic expressions in limiting cases}\label{sec:analytic}
%%%%%%%%%%%%%%%%%%%%%%%%%%%%%%%%%%%%%%

As previously mentioned, the value of $F$ can be obtained by numerical
calculations in general.
However, 
we can express it analytically in some limiting cases.  
In the following, we begin with the case where the
initial amplitude of the curvaton $\sigma_\ast$ is very small
$(\sigma_\ast \ll M_{\rm pl}$). 
In fact, this case can be further divided into two situations.
The first case is that energy density of the
curvaton once dominates the universe before the decay of the curvaton. 
The other case is that it can be always neglected compared to 
radiation during the whole history of the
universe.  These two cases correspond to $p \gg 1$ and $p \ll 1$,
respectively.  
After discussing these cases, we consider the
situation where the initial amplitude for the curvaton is large enough
to drive the second inflation. Now we look at these cases in order.

\subsubsection{The case with $\sigma_\ast \ll M_{\rm pl}$ and $p \gg1$}

Here we adopt the so-called sudden decay approximation, which can give a
good description for the case with $p \gg 1$. In this approximation, the
curvaton decays suddenly when $H=\Gamma_\sigma$ at $N=N_d$. 
That is, 
we shall regard energy densities $\rho_r$ and $\rho_\sigma$ to behave like
the step functions at the time of the curvaton decay.  
All $\rho_\sigma$ is transfered to $\rho_r$.
Hence $\rho_r$ increases by $\rho_\sigma$ at the time of the 
curvaton decay.
Then,
the equation for radiation energy density is written as
\begin{equation}
\label{eq:sudden_decay}
\frac{d \rho_r}{dN} + 4  \rho_r 
= \rho_\sigma (N_{\rm d-}) \delta (N - N_{\rm d}),
\end{equation}
where the equation is written in terms of the derivative with respect
to the number of $e$-folding and
$N_{d-}$ represents an $e$-folding number just before the time when
$H=\Gamma_\sigma$.  The coefficient on the RHS is chosen so that the
total energy density is conserved through the curvaton decay.  Then,
the equation corresponding to Eq.~\eqref{eq:def_Q} can be written as
\begin{equation}
%\label{ }
F(\sigma_\ast,m_\sigma,\Gamma_\sigma) 
= \frac{\rho_\sigma ( N_{\rm d-} ) }{\rho_r (N_{\rm d-} ) }. 
\end{equation}
Now let us introduce a new quantity $q$
defined as
\begin{equation}
\label{eq:q_def}
q \equiv 
\frac{3 \rho_\sigma (N_{\rm d-} ) }{ 4 \rho_r (N_{\rm d-} ) + 3 \rho_\sigma (N_{\rm d-} )},
\end{equation}
which roughly represents the fraction of energy density of the curvaton
to the total one at the time of the curvaton decay.
With this parameter $q$, 
$F(\sigma_\ast,m_\sigma, \Gamma_\sigma)$ can be written as
\begin{equation}
\label{eq:Q_q}
F(\sigma_\ast, m_\sigma, \Gamma_\sigma) 
=
\frac{4}{3} \frac{q}{1-q}.
\end{equation}
Since we can relate $q$ to $\sigma_\ast$, 
we can obtain the analytic expression of $Q_\sigma$ as a 
function of $\sigma_\ast, m_\sigma$ and $\Gamma_\sigma$.  
For this purpose,
let us write down the relation between $q$ and $\sigma_\ast$
in a following way.  
The Friedman equation at the time of the curvaton decay is
given by
\begin{equation}
%\label{ }
H^2 (N_{\rm d} ) = \Gamma_\sigma^2 
= \frac{1}{3 M_{\rm pl}^2} \bigg( \rho_r (N_{\rm d-}) + \rho_\sigma (N_{\rm d-}) \bigg).
\end{equation}
Denoting the $e$-folding number at the time $H=m_\sigma$ as $N_m$,
this equation can be written as
\begin{equation}
%\label{ }
\Gamma_\sigma^2 
=
\frac{1}{3M_{\rm pl}^2} 
\left(
 3 m_\sigma^2 M_{\rm pl}^2 e^{-4\bar{N}} 
 + \rho_\sigma (N_{\rm m} ) e^{-3 \bar{N}}
\right),
\end{equation}
where $\bar{N} = N_{\rm d} - N_{\rm m}$.  
Furthermore, by using the relation
\begin{equation}
%\label{ }
\frac{\rho_\sigma (N_{\rm m} )}{3 m_\sigma^2 M_{\rm pl}^2} e^{\bar{N}}
=
\frac{\rho_\sigma (N_{\rm d-} ) }{\rho_r (N_{\rm d-} )} 
= \frac{4}{3} \frac{q}{1-q},
\end{equation}
one can relate $q$ and $\rho_\sigma(N_{\rm m})$ as:
\begin{equation}
%\label{ }
\frac{\Gamma_\sigma}{m_\sigma}
= \frac{1}{16}
\left( \frac{\rho_\sigma (N_{\rm m} )}{m_\sigma^2M_{\rm pl}^2} \right)^2
\frac{(3 +q)^{1/2}}{3^{1/2}q^2} (1 -q )^{3/2}.
\end{equation}
Now we write the energy density of the curvaton at $N = N_{\rm m}$ as
$\rho_\sigma (N_{\rm m}) = (1/2) m_\sigma^2 \alpha^2 \sigma_\ast^2$
where $\alpha$ parameterizes the amplitude of the curvaton at the onset
of oscillations relative to the initial one.  The concrete value of
$\alpha$ is given in Appendix \ref{sec:alpha}.  
By using them, 
we obtain the relation between $q$ and $\sigma_\ast$ as
\begin{equation}
\label{eq:q_sigma}
\alpha^2 p
 = \frac{3^{1/4}8 q}{ (3+q)^{1/4} (1-q)^{3/4} }.
\end{equation}
By using Eqs.~\eqref{eq:Q_q} and \eqref{eq:q_sigma}, we obtain the
expressions for the derivatives of $F$ with respect to $\sigma$,
which is necessary to evaluate the inflationary parameters, 
for this limiting case as
\begin{equation}
%\label{ }
Q_\sigma = \frac{2}{3} \frac{q}{\sigma_\ast}, ~~~ 
Q_{\sigma\sigma} = \frac{2}{9} \frac{q}{\sigma_\ast^2} (3-4q-2q^2), ~~~
Q_{\sigma\sigma\sigma} = \frac{4}{27} \frac{q^2}{\sigma_\ast^3} (-18+q+20q^2+6q^3). \label{Q-sudden}
\end{equation}

In the limit $q=1$,
which is equivalent to $p\gg 1$,
$Q_\sigma$ becomes
\begin{eqnarray}
Q_\sigma = \frac{2}{3} \frac{1}{\sigma_\ast}.
\end{eqnarray}
This agrees with the result in \cite{Langlois:2004nn} where 
asymptotic form of $Q_\sigma$ for $\sigma_* \ll M_{\rm pl}$ is provided under the assumption 
that curvaton once dominates the universe before the curvaton decays.
Hence the sudden decay approximation can well describe $Q$ for
the case $p \gg 1$.

Using Eq.~(\ref{Q-sudden}),
the non-linearity parameters for $p \gg 1$ can be written as
\begin{eqnarray}
&&\frac{6}{5} f_{\rm NL}= \frac{1}{ {\left( 1+\frac{8}{9} \epsilon \frac{M_{\rm pl}^2}{\sigma_*^2} \right)}^2 } \left( 2\epsilon-\eta-\frac{32}{27} \epsilon^2 \frac{M_{\rm pl}^4}{\sigma_*^4} \right), \\
&&\tau_{\rm NL}=\frac{1}{ {\left( 1+\frac{8}{9} \epsilon \frac{M_{\rm pl}^2}{\sigma_*^2} \right)}^3 } \bigg\{ {(2\epsilon-\eta)}^2+\frac{128}{81} \epsilon^3 \frac{M_{\rm pl}^6}{\sigma_*^6} \bigg\}, \\
&&\frac{54}{25} g_{\rm NL}=\frac{1}{ {\left( 1+\frac{8}{9} \epsilon \frac{M_{\rm pl}^2}{\sigma_*^2} \right)}^3 } \left( -2\epsilon \eta-\xi^2+2\eta^2+\frac{256}{81} \epsilon^3 \frac{M_{\rm pl}^6}{\sigma_*^6} \right).
\end{eqnarray}
All the magnitudes of these non-linearity parameters become
the largest when $\sigma_* \sim \sqrt{\epsilon} M_{\rm pl}$. 
Substituting $\sigma_*=\sqrt{\epsilon} M_{\rm pl}$ to the equations above,
we find that the maximum magnitudes of $f_{\rm NL},~\tau_{\rm NL}$ 
and $g_{\rm NL}$ are all ${\cal O}(1)$.
Hence large non-Gaussianity is not produced for $p \gg 1$.

\subsubsection{The case with $\sigma_\ast \ll M_{\rm pl}$ and $p \ll1$}

Now we consider the second case where the curvaton is always subdominant
in the history of the universe.  Even in this case,
the sudden decay approximation adopted above can give a good estimate to
some extent, but its accuracy is $\mathcal{O}(10 \%)$
compared to numerical calculations\footnote{
  Using the fitting formula,
  the sudden-decay approximation can give an
  accurate estimate for some parameter range
    \cite{Gupta:2003jc}.
}.  However, another approach can give a better analytic expression
for $Q$. Detailed descriptions for the derivation of the analytic formulae
can be found in Appendix~\ref{sec:F_p}.  Here we just give the expressions.

Since here we are considering the case with $p \ll 1$, 
we expand $F$ up to the second order in $p$, 
which is necessary for obtaining the formula for the third derivative of 
$Q$ with respect to $\sigma$ because the third derivative of $p$ with
respect to $\sigma$ vanishes.
After some calculations,
we can show that the function $F$ can be expanded as

\begin{equation}
%\label{ }
F(p)=\frac{1}{6} \sqrt{\frac{\pi}{2}} \alpha^2 p  + \frac{1}{144} \alpha^4 p^2+{\cal O}(p^3),
\end{equation}
where we neglected higher order terms in $s$ assuming $s \ll 1$.
Thus the derivatives of $Q$ with respect to $\sigma$ are given for
this case as
\begin{eqnarray}
%\label{ }
&&Q_\sigma = \frac{1}{12} \sqrt{\frac{\pi}{2}} \alpha^2 
\frac{\sigma_\ast}{M_{\rm pl}^2 \sqrt{s} } \left( 1+{\cal O}(p) \right),\\
&&Q_{\sigma\sigma} = \frac{1}{12} \sqrt{\frac{\pi}{2}} \alpha^2 
\frac{1}{M_{\rm pl}^2 \sqrt{s} }\left( 1+{\cal O}(p) \right),\\
&&Q_{\sigma\sigma\sigma} = - \frac{1}{24} ( \pi -1) \alpha^4
\frac{\sigma_\ast}{M_{\rm pl}^4 s }\left( 1+{\cal O}(p) \right).
\end{eqnarray}
Here notice that, while $Q_\sigma$ and
$Q_{\sigma\sigma}$ come from the leading order term in $p$,
$Q_{\sigma\sigma\sigma}$ comes from the next-to-leading order term in $p$. 
With the expressions above, we can explicitly write down the
non-linearity parameters as:
\begin{eqnarray}
\label{eq:fnl_max}
&&\frac{6}{5} f_{\rm NL} =\frac{1}{6}  \sqrt{\frac{\pi}{2}} \alpha^2 
\frac{\epsilon}{\sqrt{s} } 
D_f \left( 
\frac{\pi \alpha^4}{144} \epsilon \frac{p}{\sqrt{s}}
\right) \left( 1+{\cal O}(p) \right), \\
\label{eq:tnl_max}
&&\tau_{\rm NL} = 
\left( \frac{1}{6}  \sqrt{\frac{\pi}{2}} \alpha^2 
\frac{\epsilon}{\sqrt{s} } \right)^2
D_\tau \left( 
\frac{\pi \alpha^4}{144} \epsilon \frac{p}{\sqrt{s}}
\right) \left( 1+{\cal O}(p) \right), \\
\label{eq:gnl_max}
&&\frac{54}{25} g_{\rm NL}=  
-(\pi-1) \sqrt{\frac{2}{\pi}} \alpha^2 
\frac{\epsilon}{\sqrt{s} } 
D_g \left( 
\frac{\pi \alpha^4}{144} \epsilon  \frac{p}{\sqrt{s}}
\right) \left( 1+{\cal O}(p) \right),
\end{eqnarray}
where the functions $D_f, D_\tau$ and $D_g$ are defined as
\begin{equation}
\label{eq:def_D}
D_f (x) \equiv \frac{x}{(1+x)^2}, ~~~
D_\tau (x) \equiv \frac{x}{(1+x)^3}, ~~~
D_g (x) \equiv \frac{x^2}{(1+x)^3}.
\end{equation}
In deriving these expressions,
we have neglected the non-Gaussianity coming from the inflaton
fluctuations because they are small quantities.
In the pure curvaton limit,
the non-linearity parameters given above become
\begin{eqnarray}
\frac{6}{5} f_{\rm NL} = \frac{24}{\sqrt{2\pi} \alpha^2} \frac{1}{p},~~~\tau_{\rm NL}=\frac{288}{\pi \alpha^4} \frac{1}{p^2},~~~\frac{54}{25}g_{\rm NL}=- \frac{288(\pi-1)}{ \sqrt{2\pi^3} \alpha^2} \frac{1}{p}.
\end{eqnarray}

Using Eqs.~(\ref{eq:fnl_max}),~(\ref{eq:tnl_max}) and (\ref{eq:gnl_max}), 
we can derive a relation among the non-linearity parameters 
specific to this case (i.e., $p \ll 1$). 
After a little arithmetic, 
we obtain
\begin{eqnarray}
-\frac{5\pi}{48 (\pi-1)} \frac{\tau_{\rm NL} g_{\rm NL}}{f_{\rm NL}^3} = 1+{\cal O} \left( p \right). \label{consistency}
\end{eqnarray}
When $p \ll 1$,
$p$ can be written using $f_{\rm NL}$ and $\tau_{\rm NL}$ as
\begin{equation}
%\label{ }
p= \sqrt{ \frac{\pi}{2}} \frac{5184}{125 \alpha^2} f_{\rm NL}^{-1} {\left( \frac{f_{\rm NL}^2}{ \tau_{\rm NL}} \right)}^2= {\cal O} \left( f_{\rm NL}^{-1} {\left( \frac{f_{\rm NL}^2}{ \tau_{\rm NL}} \right)}^2 \right).
\end{equation}
Because of the inequality Eq.~\eqref{eq:inequality},
the magnitude of $p$ is smaller than that of $f_{\rm NL}^{-1}$.
Hence for the case of the very large non-Gaussianity ($f_{\rm NL}\gg 1$)
which occurs only when $p \ll 1$,
the RHS in Eq.~(\ref{consistency}) becomes very close to unity and the equation
provides a simple consistency relation between the bispectrum and the trispectrum.
Note that this relation holds not only for the pure curvaton model
but also for the mixed model of the inflaton and the curvaton. 
Neglecting the second term on the RHS in Eq.~(\ref{consistency}),
we can further derive the inequality for $f_{\rm NL}$ and $g_{\rm NL}$ 
as, assuming $f_{\rm NL}$ being positive, 
\begin{equation}
%\label{ }
g_{\rm NL} \le -\frac{20 (\pi-1)}{3\pi} f_{\rm NL}.
\end{equation}
Thus if the very large non-Gaussianity is detected in the future, 
we can discriminate the mixed model of the inflaton and the curvaton
from other scenarios that also generate large non-Gaussianity
by using these consistency relations.

Furthermore, notice that the argument $x$ of the functions $D_f, D_\tau$ and $D_g$ 
corresponds to the ratio of the inflaton's fluctuation to the curvaton's, i.e., 
$x = 2 \epsilon M_{\rm pl}^2 Q_\sigma^2=\zeta_{\rm cur}^2 / \zeta_{\rm inf}^2$ 
where $\zeta_{\rm inf}$ and $\zeta_{\rm cur}$
are the curvature fluctuations generated from the inflaton and the curvaton respectively.
Since all these functions take the maximum values at $x\sim 1$,
non-linearity parameters become the largest when 
the fluctuations from the curvaton are comparable to those generated
from the inflaton with $\epsilon$ and $s$ being fixed.
In this case, the initial amplitude of the curvaton 
becomes
$\sigma_\ast \sim \sigma_{\ast, \rm max} \sim \sqrt{s/\epsilon} M_{\rm pl}$.
If $\sigma_\ast$ is smaller than $\sigma_{\ast, \rm max}$,
the curvature perturbations are dominated by the inflaton fluctuations
which are highly Gaussian.
Hence when $\sigma_\ast$ is smaller than $\sigma_{\ast, \rm max}$,
the magnitudes of the non-linearity parameters take smaller values.
On the other hand,
if $\sigma_\ast$ is larger than $\sigma_{\ast, \rm max}$,
it becomes similar to the pure curvaton case, in which 
fluctuations from the curvaton dominate over those of the inflaton. 
In this case, $f_{\rm NL}$ and $g_{\rm NL}$ are proportional to $1/\sigma_\ast^2$
and $\tau_{\rm NL}$ are proportional to $1/\sigma_\ast^4$.
Hence, as we take the value of $\sigma_\ast$ be larger than $\sigma_{\ast, \rm max}$,
the magnitude of the non-linearity parameters becomes smaller.
Thus if we vary $\sigma_*$ under the condition that $\epsilon$ and $s$ are fixed,
in which we are changing the amplitude of $\zeta_{\rm cur}$ with $\zeta_{\rm inf}$ being fixed,
the non-Gaussianity becomes the largest when $\sigma_\ast \sim \sigma_{\ast, \rm max}$,
i.e., 
when the contributions from the inflaton and the curvaton are comparable.
On the other hand,
if we instead vary $\epsilon$ under the condition that $s$ and $\sigma_*$ are fixed,
where we are changing the amplitude of $\zeta_{\rm inf}$ with $\zeta_{\rm cur}$ being fixed,
the non-Gaussianity becomes the largest when the contributions from
the inflaton are absent.

The maximum values of $f_{\rm NL}, \tau_{\rm NL}$ and the minimum value
of $g_{\rm NL}$ for the variation of $\sigma_*$ are given by
\begin{eqnarray}
%\label{ }
&&\frac{6}{5}f_{\rm NL, max}=\frac{1}{24} \sqrt{\frac{\pi}{2}}\alpha^2 \frac{\epsilon}{\sqrt{s}} \sim \frac{1}{p_{\rm max}},\\
&&\tau_{\rm NL, max} =\frac{\pi \alpha^4}{486} \frac{\epsilon^2}{s}\sim \frac{1}{p_{\rm max}^2}, \\
&&\frac{54}{25}g_{\rm NL, min} = -\frac{4(\pi-1)}{27} \sqrt{\frac{\pi}{2}} \alpha^2 \frac{\epsilon}{\sqrt{s}}\sim \frac{1}{p_{\rm max}},
\end{eqnarray}
where $p_{\rm max}$ is the value of $p$ with $\sigma_\ast = 
\sigma_{\ast,\rm max}$, which 
is much smaller than $1$.
We find that the magnitudes of these non-linearity parameters 
have the same $p$ dependence as
those in the case of the pure curvaton, i.e.,
$f_{\rm NL}$ and $g_{\rm NL}$ are roughly given by the inverse of the
fraction of the curvaton energy density to the total one at the time
of the curvaton decay and $\tau_{\rm NL}$ is roughly given by $f_{\rm NL}^2$\footnote{
In Ref.~\cite{Huang:2008ze},
by requiring that the homogeneous energy density of the
curvaton should be larger than its gradient energy density
and the decay rate $\Gamma_\sigma$ is at least of order that given by
the gravitational strength, an 
upper bound on $f_{\rm NL}$ 
was provided as a function of the tensor-to-scalar ratio in the pure curvaton case.
}.

%The fact that non-linearity parameters become largest in magnitude
%when $x \sim 1$ seems to contradict with a naive expectation that
%the non-Gaussianity always decreases when we add $\zeta_{\rm inf}$
%to the total curvature fluctuations and that the non-Gaussianity becomes
%the largest when the contribution from the curvaton is much larger than 
%that from the inflaton.
%This apparent paradox is due to the difference of the variables we change.
%Indeed,
%if we decrease $\epsilon$ with $\sigma_*$ and $s$ being fixed,
%which is equivalent to the increase of $\zeta_{\rm inf}$ with 
%$\zeta_{\rm cur}$ being fixed,
%magnitude of the non-linear parameters always decreases.

As a final remark of this subsection, we comment on the relation 
$\tau_{\rm NL}\sim f_{\rm NL}^2$ which holds 
for $\sigma_\ast=\sigma_{\ast, \rm max}$.
In fact, this relation breaks down when $\sigma_\ast$ deviates
from $\sigma_{\ast, \rm max}$.
From Eq.~(\ref{eq:def_D}),
we have
\begin{eqnarray}
\frac{D_\tau (x)}{D_f^2(x)}=1+\frac{1}{x}.
\end{eqnarray}
Hence when the curvaton contribution to the curvature perturbations
is subdominant,
i.e.,
$x \ll 1$,
$\tau_{\rm NL}$ is enhanced by a factor of $x^{-1}$ compared with $f_{\rm NL}^2$.
This shows that the trispectrum as well as the bispectrum could also 
be important in detecting the non-Gaussianity of the primordial perturbations.
As we will see in Section \ref{sec:obs}
where $f_{\rm NL}, g_{\rm NL}$ and $\tau_{\rm NL}$ are calculated
for various inflation models in the mixed inflaton and curvaton scenario,
there are parameter regions where $f_{\rm NL}, g_{\rm NL} \lesssim {\cal O}(1)$
but $\tau_{\rm NL} \gg 1$.
In such a case,
the leading non-Gaussianity comes not from the bispectrum but 
from the trispectrum through the $\tau_{\rm NL}$ terms.

\subsubsection{The case with $\sigma_\ast \gg M_{\rm pl}$}

When the initial amplitude is large enough, the curvaton field can drive
the second inflation after the first inflation caused by the
inflaton. 
In this case, $N_{\rm R}$ is calculated from Eq.~(\ref{eq:N_cur2}) 
and is given by
\begin{equation}
%\label{ }
N_{\rm R} \simeq \frac{1}{4 M_{\rm pl}^2} (\sigma_\ast^2 - \sigma_e^2 )+C. \label{sigma_large}
\end{equation}
Thus, in this case, $Q_\sigma,~Q_{\sigma \sigma}$ and $Q_{\sigma \sigma \sigma}$ become
\begin{equation}
%\label{ }
Q_\sigma = \frac{1}{2} \frac{\sigma_\ast}{M_{\rm pl}^2},~~~
Q_{\sigma\sigma} = \frac{1}{2M_{\rm pl}^2} ,~~~
Q_{\sigma\sigma\sigma} = 0.
\end{equation}
Then,
the non-linearity parameters are given by
\begin{eqnarray}
&&\frac{6}{5}f_{\rm NL}=\frac{1}{ {\left( 1+\frac{1}{2}\epsilon \frac{\sigma_*^2}{M_{\rm pl}^2} \right)}^2 } \left( 2\epsilon-\eta+\frac{1}{2}\epsilon^2 \frac{\sigma_*^2}{M_{\rm pl}^2} \right), \label{eqf}\\
&&\tau_{\rm NL}=\frac{1}{ {\left( 1+\frac{1}{2}\epsilon \frac{\sigma_*^2}{M_{\rm pl}^2} \right)}^3 } \bigg\{ {(2\epsilon-\eta)}^2+\frac{1}{2}\epsilon^3 \frac{\sigma_*^2}{M_{\rm pl}^2} \bigg\}, \label{eqt}\\
&&\frac{54}{25} g_{\rm NL}=\frac{1}{ {\left( 1+\frac{1}{2}\epsilon \frac{\sigma_*^2}{M_{\rm pl}^2} \right)}^3 } (-2\epsilon \eta-\xi^2+2\eta^2).
\end{eqnarray}
As for $f_{\rm NL}$ and $\tau_{\rm NL}$,
the contributions from the curvaton become the largest
when $\sigma_* \sim M_{\rm pl}/\sqrt{\epsilon}$.
Substituting $\sigma_* \sim M_{\rm pl}/\sqrt{\epsilon}$ to
Eqs.~(\ref{eqf}) and (\ref{eqt}),
we find that the maximum contributions of the curvaton to
$f_{\rm NL}$ and $\tau_{\rm NL}$ are ${\cal O}(\epsilon)$
and ${\cal O}(\epsilon^2)$,
respectively.
As for $g_{\rm NL}$,
addition of the curvaton always reduces the magnitude of $g_{\rm NL}$.
Hence $f_{\rm NL}, \tau_{\rm NL}$ and $g_{\rm NL}$ are
at most ${\cal O}(\epsilon, \eta), {\cal O}(\epsilon^2, \epsilon \eta, \eta^2)$
and ${\cal O}(\epsilon^2, \epsilon \eta, \eta^2, \xi^2)$,
respectively.

%%%%%%%%%%%%%%%%%%%%%%%%%%%%%%%
\subsection{Corresponding Scales} \label{sec:scale}
%%%%%%%%%%%%%%%%%%%%%%%%%%%%%%%

To compare the prediction for the primordial curvature fluctuations and
non-Gaussianity with observations, we need to specify when the present
cosmological scale exited the horizon during inflation. Since
$k_\ast = a_\ast H_\ast$ holds when the scale with the wave number
$k_\ast$ crossed the horizon, the reference scale $k_{\rm ref}$ where we
probe the primordial fluctuations at the present time is related to that
at the horizon crossing during inflation as
\begin{equation}
%\label{ }
\frac{k_{\rm ref}}{a_0 H_0} = \frac{a_\ast H_\ast}{a_0 H_0},
\end{equation}
where $a_0$ and $H_0$ are the scale factor and the Hubble parameter at
present. The ratio of $a_\ast$ to $a_0$ in the RHS can be 
divided into several parts as,
\begin{equation}
%\label{ }
\frac{k_{\rm ref}}{a_0 H_0} = 
\frac{a_\ast}{a_{\rm end}} 
\frac{a_{\rm end}}{a_{\phi \rm reh}}
\frac{a_{\phi \rm reh}}{a_f}
\frac{a_f}{a_0}
\frac{H_\ast}{H_0}.
\end{equation}
Here $a_{\rm end}, a_{\phi \rm reh}$ and $a_f$ are the scale factors at the
times when the inflation driven by the inflaton ends, the inflaton decays
to reheat the universe, and some time after the curvaton decays into
radiation, respectively. For the definiteness, we take $ H \simeq
10^{-2} \Gamma_\sigma$ as the time corresponding to $a_f$ so that
Eq.~\eqref{eq:N_cur} gives a good description.  By taking logarithm of
both sides and reminding that the number of $e$-folding during inflation
is $N_{\rm inf} = \log (a_\ast / a_{\rm end})$, $N_{\rm inf}$ can be
written as
\begin{equation}
\label{eq:N_e}
N_{\rm inf}
= 
- \log \frac{k_{\rm ref}}{a_0 H_0} 
+ \log \frac{a_{\rm end}}{a_{\phi \rm reh}}
+ \log \frac{a_{\phi\rm reh}}{a_f}
+ \log \frac{a_f}{a_0} 
+ \log \frac{H_\ast}{H_0}.
\end{equation}
For the first term, we take $k_{\rm ref} = 0.002~{\rm Mpc}^{-1}$ in the
analysis.  The second term in the RHS corresponds to $-N_d$ in
Eq.~(\ref{eq:N_tot}), i.e., the $e$-folding number from the end of
inflation to the decay of the inflaton to the radiation.  
When the potential of the inflaton near the minimum is written as 
$V \propto \phi^\alpha$, 
its energy density decreases as $\rho_\phi \propto a^{-6\alpha/(\alpha +2)}$ 
\cite{Turner:1983he}.  
Thus the second term can be written as
\begin{equation}
%\label{ }
\log \frac{a_{\rm end}}{a_{\phi \rm reh}} 
=
\frac{\alpha + 2}{6\alpha} \log \frac{\rho_{\phi \rm reh}}{\rho_{\rm end}}  
\simeq 
\frac{\alpha + 2}{3\alpha} \log \frac{\Gamma_\phi}{H_{\rm end}},  
\end{equation}
where $\rho_{\rm end}$ and $\rho_{\phi \rm reh}$ are energy density of
inflaton when inflation ends and that of radiation at reheating.  For
the second approximate equality, we use the sudden decay approximation
for the reheating $\Gamma_\phi = H (t_{\phi \rm reh})$ with $\Gamma_\phi$
being the decay rate of the inflaton.
For the analysis in the subsequent section,
we assume $\Gamma_\phi=10^{-2} H_{\rm end}$ for definiteness.

The third term corresponds to $-N_R$ in Eq.~(\ref{eq:N_tot}).
The effect of the curvaton on the total $e$-folding number is contained
in this term (see Eq.~(\ref{eq:N_cur})).

As for the fourth term,
we assume that no more entropy is produced after the curvaton decays. 
Thus this term can be rewritten by using the conservation of the entropy
density per comoving volume. 
Since the entropy density is given by 
$s = (2 \pi / 45) g_{\ast s}T^3$ with $g_{\ast s}$ being the 
total number of effective massless degrees of freedom,
we have the following relation,
\begin{equation}
%\label{ }
 \log \frac{a_f}{a_0} 
= \frac{1}{3} \log \frac{s_0}{s_f}=\frac{1}{3} \log \frac{g_{\ast s 0} T_0^3}{g_{\ast s f} T_f^3}.
\end{equation}
For $g_{\ast s f}$ at the time of $a_f$, 
we take $ g_{\ast s f}=100$.

%%%%%%%%%%%%%%%%%%%%%%%%%%%%%%%
\section{Observational quantities in the mixed models with the inflaton and the curvaton} \label{sec:obs}
%%%%%%%%%%%%%%%%%%%%%%%%%%%%%%%

Now we discuss primordial curvature fluctuations and non-Gaussianity in
the mixed models with the inflaton and the curvaton adopting some
concrete inflation models.  Once the potential for the inflaton is
specified, using the formalism developed above, we can make firm
predictions for the scalar spectral index, the tensor-to-scalar ratio
and the non-Gaussianity for the model.  
The issues of the modification to the spectral index, 
its running and tensor modes have been studied in the
case where the energy density of the curvaton should dominate the
universe at late time \cite{Langlois:2004nn,Moroi:2005kz,Moroi:2005np}.
Here we also include the case where the curvaton is always subdominant
in the whole history of the universe. 
Furthermore, we give the predictions for the non-linearity parameters 
$f_{\rm NL}, \tau_{\rm NL}$ and $g_{\rm NL}$ for inflation models
in the scenario.  
In the following we consider chaotic inflation, 
new inflation, and hybrid inflation in order.

\subsection{Chaotic inflation}

\begin{figure}[htb]
\begin{center}
\scalebox{1}{
\includegraphics{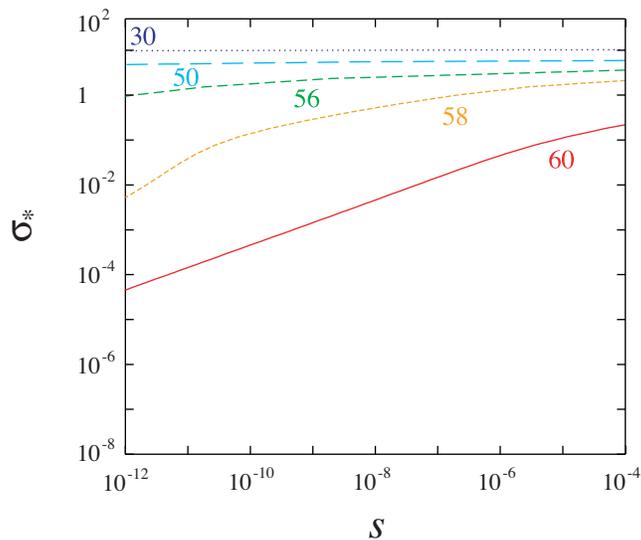}
}
\caption{Contours of the $e$-folding number $N$ in the $s$--$\sigma_\ast$ 
plane. Here we assumed the chaotic inflation model with $n=4$ for 
concreteness.
$\sigma_*$ is shown in units of $M_{\rm pl}$.
  }
\label{fig:chaotic4_efold}
\end{center}
\end{figure}

\begin{figure}[htb]
\begin{center}
\scalebox{1}{
\includegraphics{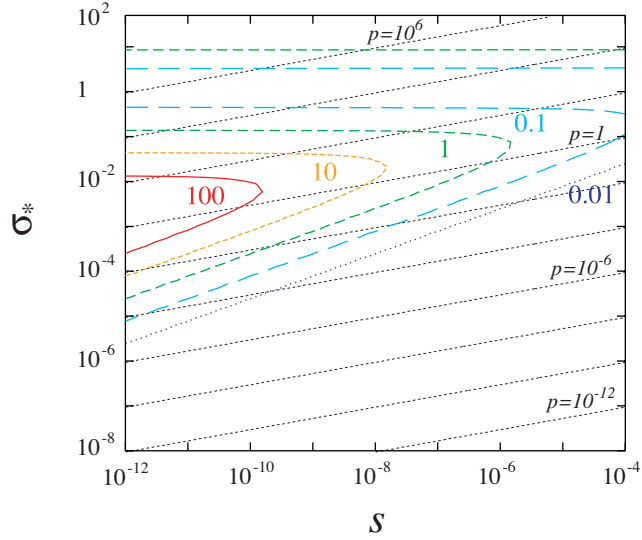}
}
\caption{Contours of the ratio of the contribution from
the curvaton fluctuation and the inflaton fluctuation 
$\zeta_{\rm cur}^2/ \zeta_{\rm inf}^2$ are shown in the $s$--$\sigma_\ast$ 
plane. Here we assumed the chaotic inflation model with $n=4$ for 
concreteness.
For reference, 
we also show contours of $p$ which enables us to see the region where 
an analytic expression is valid provided in the previous section.
  }
\label{fig:chaotic4_ratio}
\end{center}
\end{figure}

\begin{figure}[htb]
\begin{center}
\scalebox{0.8}{
\includegraphics{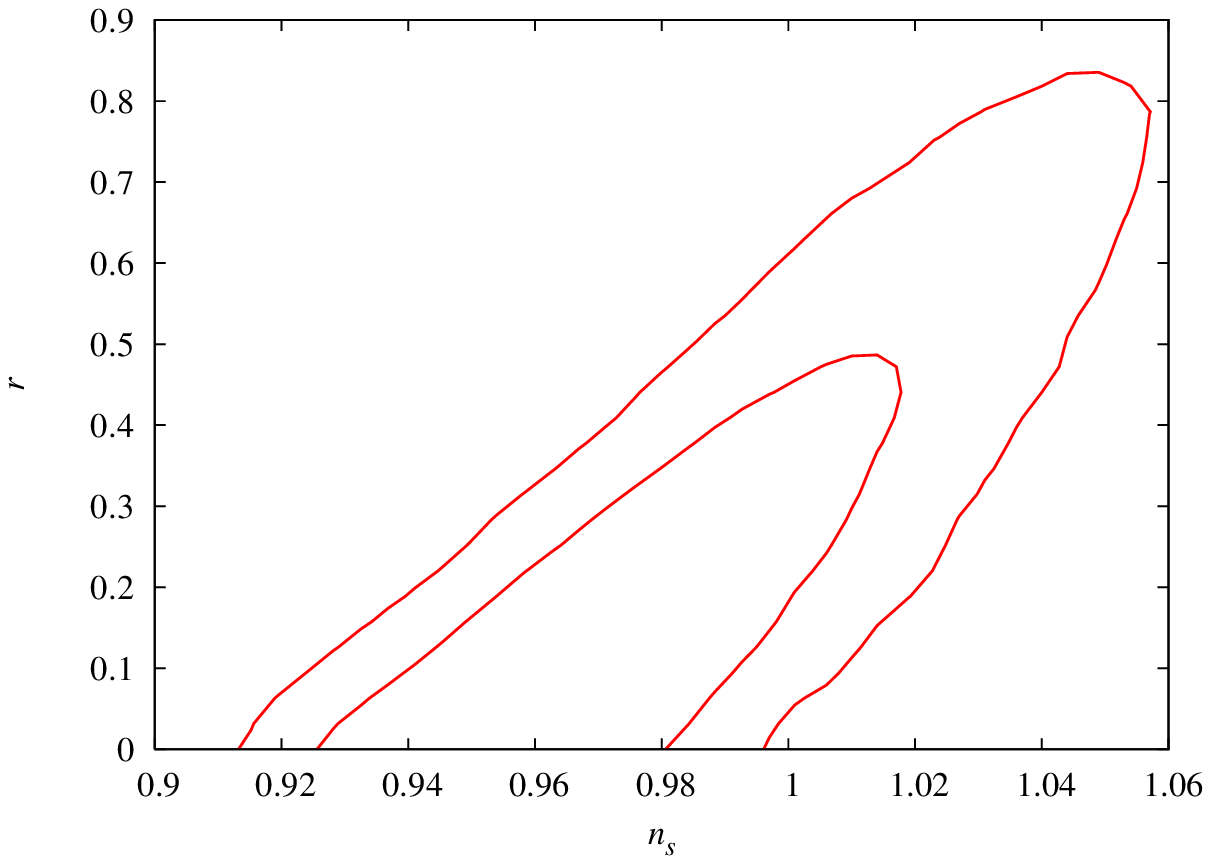}
}
\caption{1$\sigma$ and 2$\sigma$ allowed contours from WMAP3 alone are shown,
which is generated from the chain provided in the webpage \cite{LAMBDA}.
  }
\label{fig:wmap3}
\end{center}
\end{figure}

Chaotic inflation \cite{Linde:1983gd} is described by the potential of
the form,\footnote{This type of a simple polynomial potential can be
realized in supergravity
\cite{Goncharov:1983mw,Goncharov:1985ka,Murayama:1993xu,Kawasaki:2000yn,Kawasaki:2000ws,Kadota:2007nc,Kawano:2007gg,Kadota:2008pm}.}
\begin{equation}
\label{eq:V_chaotic}
V =  \lambda M_{\rm pl}^4 \left( \frac{\phi}{M_{\rm pl}} \right)^{n},
\end{equation}
where $\lambda$ is a model parameter which can be fixed by requiring
that the fluctuations have the right amplitude to be consistent with
observations. 
To fix the model parameters, 
we impose the WMAP normalization \cite{Liddle:2006ev} for the primordial curvature
fluctuations given in Eq.~\eqref{eq:P_zeta2} to fix the value of
$\lambda$. 
For different models of inflation discussed in the following,
we also use the WMAP normalization to fix one of the parameters in the
inflaton potential.

Here we consider some versions of chaotic inflation by taking
several values of $n$ which is assumed to be positive even number.
With the potential Eq.~\eqref{eq:V_chaotic}, the slow-roll parameters in
this model are given by
\begin{equation}
%\label{ }
\epsilon = \frac{1}{2}n^2 \left( \frac{M_{\rm pl}}{\phi_\ast} \right)^2, ~~~
\eta = n ( n - 1) \left( \frac{M_{\rm pl}}{\phi_\ast} \right)^2, ~~~
\xi^2 = n (n -1) (n -2) \left( \frac{M_{\rm pl}}{\phi_\ast} \right)^4.
\end{equation}
The number of $e$-folding during inflation can be written as 
\begin{equation}
\label{eq:N_chaotic}
N_{\rm inf} = \frac{1}{2 n M_{\rm pl}^2}  ( \phi_\ast^2 - \phi_{\rm end} )^2,
\end{equation}
where $\phi_{\rm end}$ refers to the value of the inflaton at the end of
inflation. In general, we can assume $\phi_\ast \gg \phi_{\rm end}$. 
Thus we have a simple relation between $\phi_\ast$ and $N_{\rm inf}$.  When
fluctuations of the inflaton alone are assumed to exist, the spectral index, its
running and the tensor-to-scalar ratio are given by, using the number of
$e$-folding $N_{\rm inf}$,
\begin{eqnarray}
\label{eq:ns_chaotic}
n_s -1 &=& - \frac{n +2}{2N_{\rm inf}}, \\
\label{eq:nrun_chaotic}
n_{\rm run} &=&  - \frac{n + 2}{2N_{\rm inf}^2}, \\
\label{eq:r_chaotic}
r &=&  \frac{4 n}{N_{\rm inf}}.
\end{eqnarray}
When fluctuations from the inflaton alone are considered, 
the non-linearity parameters are very small, which are
of the order of slow-roll parameters.  An
explicit expression can be obtained by setting the curvaton contributions
such as $Q_\sigma$ and $Q_{\sigma\sigma}$ to zero in Eqs.~\eqref{eq:f_NL},
\eqref{eq:tau_NL} and \eqref{eq:g_NL}.

In the following, we consider 
the case with $n=2,4$ and $6$, 
then discuss how the curvaton fluctuations change
the predictions of the original inflation models.  As we discussed in
the previous section, the contribution from the curvaton fluctuations
modifies $n_s, n_{\rm run}$ and $r$ as Eqs.~\eqref{eq:n_s},
\eqref{eq:n_run} and \eqref{eq:r}.  
In addition, the curvaton affects the background evolution, 
which changes the number of $e$-folding
during inflation.  When the oscillating curvaton field dominates, its
energy density decreases as $\rho_\sigma \propto a^{-3}$. If this phase
lasts long, it reduces $N_{\rm inf}$.  When the curvaton drives the second
inflation, the number of $e$-folding is drastically reduced as 
$N_{\rm inf} \sim 20$--$30$. In this case, as can be read off from Eqs.~\eqref{eq:ns_chaotic},
\eqref{eq:nrun_chaotic} and \eqref{eq:r_chaotic}, the spectral index
becomes more red-tilted, its running becomes larger and the
tensor-to-scalar ratio is also a bit larger due to the change of $N_{\rm inf}$.
Thus when one considers the effects of the curvaton, the modification to
the background evolution is also important. 

To see how the curvaton
parameters affect the $e$-folding number during inflation, 
contours of $N_{\rm inf}$ in the $s$--$\sigma_\ast$ plane are shown 
in Fig.~\ref{fig:chaotic4_efold}\footnote{
For the range of $s$,  we take $ 10^{-12} \le s \le 10^{-4}$  
in the analysis here. Although $s$ is 
determined by the combination of the decay rate the curvaton 
mass as $s= \Gamma_\sigma / m_\sigma$, 
to make our analysis general, we do not specify the values of $\Gamma_\sigma$ 
and $m_\sigma$. 
However, one should keep in mind that there are some
constraints on $\Gamma_\sigma$ coming from the bounds
on the reheating temperature $T_{\rm reh}$.
A possible upper bound for $T_{\rm reh}$ comes from 
the consideration of the gravitino problem. To avoid the overproduction of gravitinos, 
the reheating temperature should be low enough as $T_{\rm reh} \sim 10^6$ GeV
\cite{Kawasaki:2004qu,Kawasaki:2004yh}. 
In addition, $T_{\rm reh} \sim 10$ MeV is a possible lower bound 
not to spoil the success of BBN. 
(Detailed discussion on this issue can be found in  \cite{Kawasaki:1999na, Kawasaki:2000en, Hannestad:2004px, Ichikawa:2005vw, Ichikawa:2006vm}.)
}.  
For concreteness, 
we assumed $n=4$ to plot the figure. 
In the figure,
$\sigma_*$ is shown in units of $M_{\rm pl}$.

As is clearly seen from the figure, 
when the initial amplitude is large, 
$N_{\rm inf}$ is significantly reduced due to the second inflation driven by
the curvaton.  
On the other hand,
when $\sigma_* \ll M_{\rm pl}$,
dependence of $N_{\rm inf}$ on $\sigma_*$ and $s$ becomes mild.
The qualitative behavior of this result can be understood
as follows.
There are two possible sources that affect $N_{\rm inf}$ due to the
existence of the curvaton.
The first one is that the curvaton adds extra
$e$-folding number $Q$ compared to when the curvaton is 
absent.
As a result,
$N_{\rm inf}$ must be replaced with $N_{\rm inf}-Q$. 
The second one is that if the curvaton fluctuations
contribute to the density fluctuations,
the energy scale of inflation must decrease so that the sum of 
the fluctuations originating from both the inflaton and
the curvaton satisfies the WMAP normalization.
This effect is described by $Q_\sigma$.
Hence $N_{\rm inf}$ is a function of $Q$ and $Q_\sigma$,
i.e.,
$N_{\rm inf}=N_{\rm inf}(Q,Q_\sigma)$.

When $\sigma_* \gg M_{\rm pl}$,
$Q$ depends only on $\sigma_*$ (see Eq.~(\ref{sigma_large})).
Hence the slope of the contour in Fig.~\ref{fig:chaotic4_efold}
becomes zero.
On the other hand,
when $\sigma_* \ll M_{\rm pl}$,
$Q$ becomes a function of $p$ while $Q_\sigma$ becomes
a function of $p$ and $\sigma_*/\sqrt{s}$.
Hence we can write $N_{\rm inf}$ as $N_{\rm inf}=N_{\rm inf}(p,\sigma_*/\sqrt{s})$.
When $p \ll 1$,
the modification of $N_{\rm inf}$ due to the existence of the
curvaton mainly comes from $Q_\sigma$ which is a function
of $\sigma_*/\sqrt{s}$ at the leading order in $p$
because $Q$ is very small ($Q={\cal O}(p)$).
Hence for $p \ll 1$,
the slope of the contour approaches $1/4$.
When $p \gg 1$,
both $Q$ and $Q_\sigma$ can be important and the slope
can vary along the contour.

Furthermore,
to see in what range of $s$ and $\sigma_\ast$ 
fluctuations from the curvaton dominates over
that from the inflaton, 
the contours of the ratio of the squared contributions from 
the curvaton and the inflaton, 
$\zeta_{\rm cur}^2/ \zeta_{\rm inf}^2 = 2 M_{\rm pl}^2 \epsilon Q_\sigma^2$,
are shown in the $s$--$\sigma_\ast$ plane in Fig.~\ref{fig:chaotic4_ratio}. 
For concreteness, we again assumed the chaotic inflation model with $n=4$. 
We also plotted contours of $p=\sigma_*^2/(M_{\rm pl}^2 \sqrt{s})$,
which enable us to see in what region the analytic estimate
we discussed in the previous section can be valid.

\begin{figure}[h]
\begin{center}
\scalebox{1}{
\includegraphics{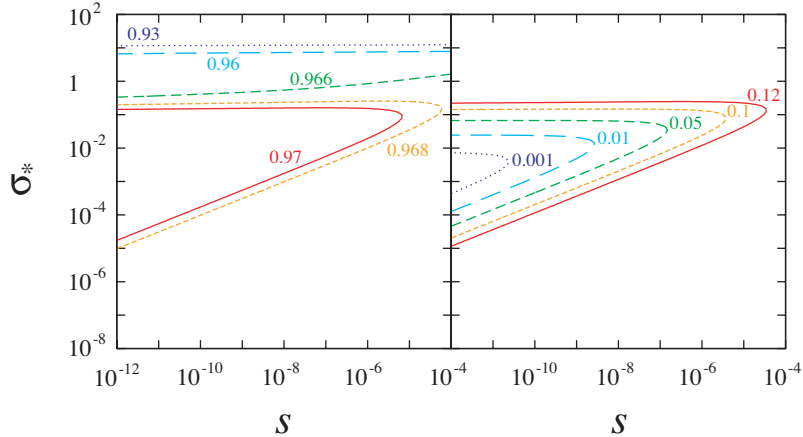} 
}
\caption{Contours of $n_s$ (left panel) and $r$ (right panel) 
 in the $s$--$\sigma_\ast$ 
plane for the chaotic inflation model with $n=2$.
In the absence of the curvaton contribution, the spectral index and 
the tensor-to-scalar ratio are $n_s=0.967$ and $r=0.133$ for $N_{\rm inf}=60.3$.
  }
\label{fig:chaotic2_ns}
\end{center}
\end{figure}
\begin{figure}[h]
\begin{center}
\scalebox{1}{
\includegraphics{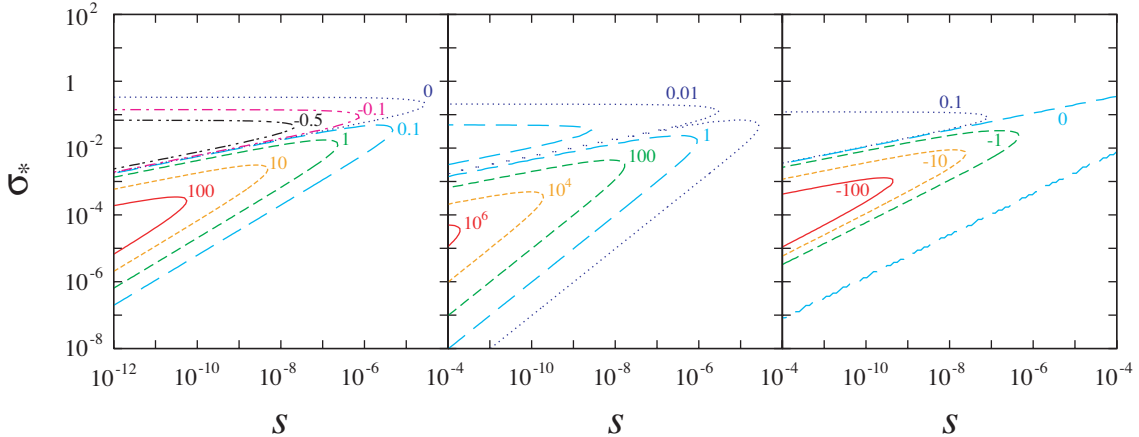}
}
\caption{Contours of $f_{\rm NL}$ (left panel), $\tau_{\rm NL}$ (center panel) and 
$g_{\rm NL}$ (right panel) in the $s$--$\sigma_\ast$ 
plane for the chaotic inflation model with $n=2$.
  }
\label{fig:chaotic2_fnl}
\end{center}
\end{figure}

\begin{figure}[h]
\begin{center}
\scalebox{1}{
\includegraphics{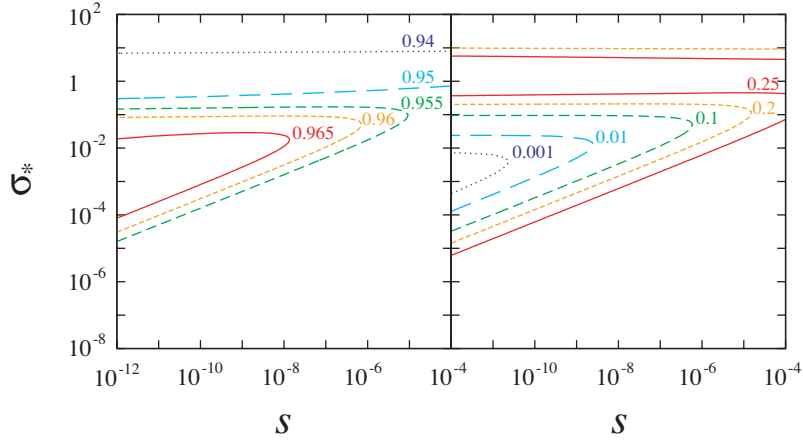} 
}
\caption{Contours of $n_s$ (left panel) and $r$ (right panel) 
   in the $s$--$\sigma_\ast$ plane for the
  chaotic inflation model with $n=4$. 
 In the absence of the curvaton contribution,
$n_s=0.950$ and $r=0.265$ for $N_{\rm inf}=60.3$. }
\label{fig:chaotic4_ns}
\end{center}
\end{figure}
\begin{figure}[h]
\scalebox{1}{
\includegraphics{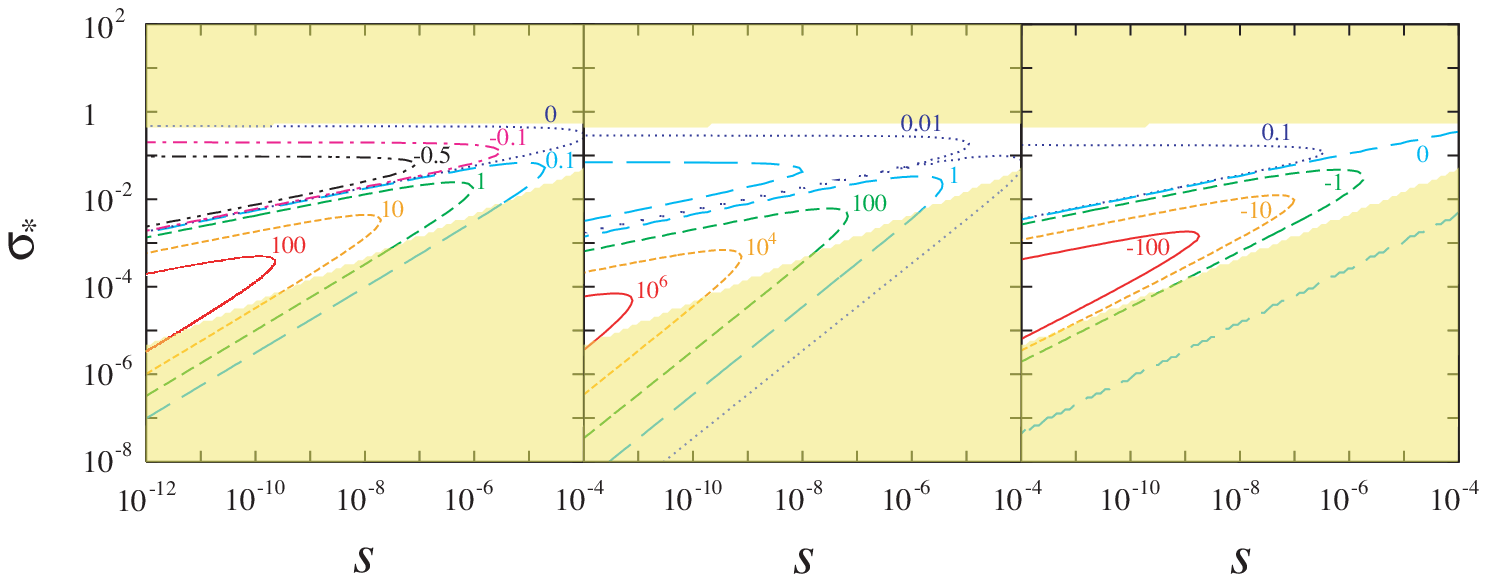} 
}
\caption{Contours of $f_{\rm NL}$ (left panel), $\tau_{\rm NL}$
  (center panel) and $g_{\rm NL}$ (right panel) in the
  $s$--$\sigma_\ast$ plane for the chaotic inflation model with $n=4$.
  The shaded region is excluded by WMAP3 data. }
\label{fig:chaotic4_fnl}
\end{figure}
\begin{figure}[htb]
\begin{center}
\scalebox{1}{
\includegraphics{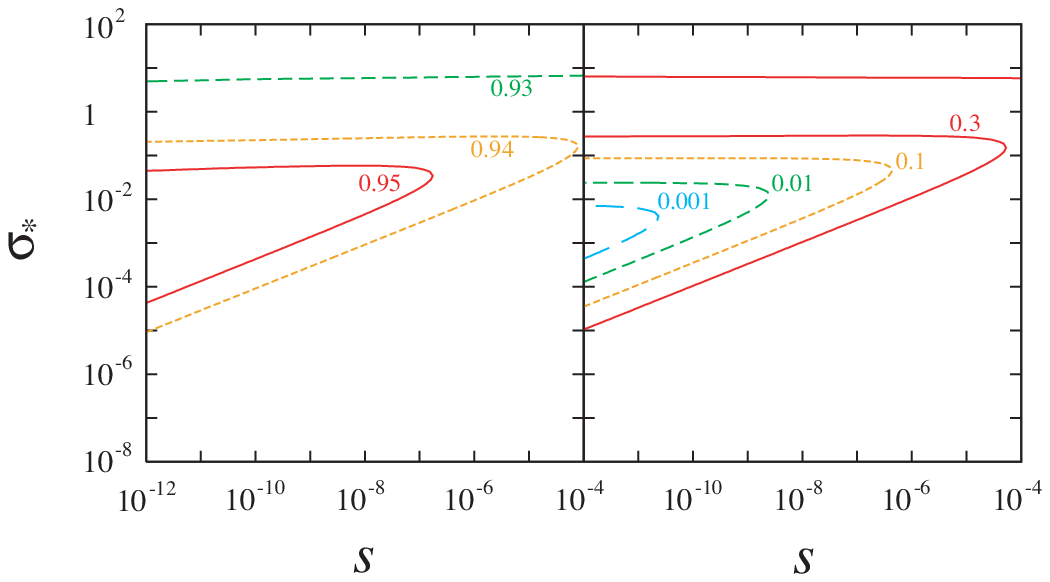}
}
\caption{Contours of $n_s$ (left panel) and $r$ (right panel)
in the $s$--$\sigma_\ast$ plane for the chaotic inflation model with $n=6$. 
 In the absence of the curvaton contribution, the spectral index and 
the tensor-to-scalar ratio are $n_s=0.937$ and $r=0.375$ for $N_{\rm inf}=64.0$. }
\label{fig:chaotic6_ns}
\end{center}
\end{figure}
\begin{figure}[h]
\scalebox{1}{
\includegraphics{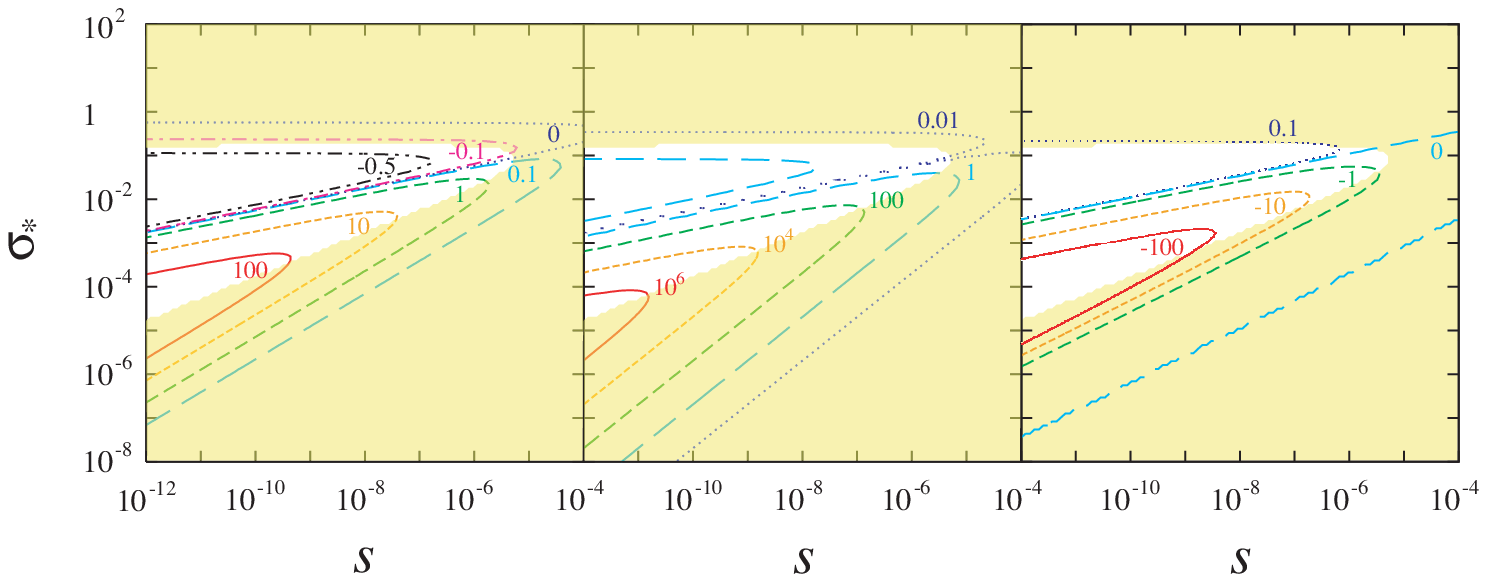}
}
\caption{Contours of $f_{\rm NL}$ (left panel), $\tau_{\rm NL}$
  (center panel) and $g_{\rm NL}$ (right panel) in the
  $s$--$\sigma_\ast$ plane for the chaotic inflation model with $n=6$.
  The shaded region is excluded by WMAP3 data. }
\label{fig:chaotic6_fnl}
\end{figure}

Now we discuss the inflationary parameters such as $n_s, r$
and the non-linearity parameters in each case. First we study
the case with $n=2$. In this case, if we assume $N_{\rm inf}=50$ and $60$, the
spectral index and the tensor-to-scalar ratio become 
$(n_s, r) =(0.96,0.16)$ and $(n_s, r) = (0.97,0.13)$, respectively. 
The running of the spectral index is negligible for this model.
To compare
these values with observations, we show 1 and 2$\sigma$ allowed regions
from WMAP3 alone analysis in the $n_s$--$r$ plane in
Fig.~\ref{fig:wmap3}.  By looking at the figure, we can see that the
case with $n=2$ is favored by WMAP3 data regardless of the 
curvaton contribution.  

If we include the curvaton,
then the non-linearity parameters will become large and
$n_S$ and $r$ will also change
\footnote{
Even when fluctuation of the curvaton is large enough to affect
the values of these quantities, the running of the spectral index remains
negligible in this model.}.
Hence it is interesting to check that the inclusion of the 
curvaton does not spoil the successful values of $n_s$ and $r$,
keeping the non-linearity parameters large.

In Fig.~\ref{fig:chaotic2_ns}, contours of $n_s$
and $r$ are shown for the case with $n=2$ when the curvaton is included
in the $s$--$\sigma_\ast$ plane. 
With our assumptions
explained in the previous section, the number of $e$-folding becomes as 
$N_{\rm inf} =60.3$, which yields the spectral index and the tensor-to-scalar ratio as 
$(n_s, r) = (0.967,0.133)$ without the curvaton contribution. We can see that,
even when the curvaton is introduced to affect the primordial
fluctuation, it is consistent with current observations.  In
Fig.~\ref{fig:chaotic2_fnl}, the non-linearity parameters are shown also in
the $s$--$\sigma_\ast$ plane.  When the curvaton drives the second
inflation, the sign of $f_{\rm NL}$ becomes negative and its size is
$\mathcal{O}(1)$ in the magnitude. When the curvaton dominates at least
once in the history of the universe but there is no
second inflation which corresponds to the region where $\sigma_\ast$ is
small and $s$ is large, the non-linearity parameters become
positive but the size is also $\mathcal{O}(1)$.  As discussed in the previous
section, non-Gaussianity becomes large when $p \ll 1$, which can be
seen from the figure.  This is why the long axis of contour ellipses
lies along the line with $p = \sigma_\ast^2 / M_{\rm pl}^2 \sqrt{s}$ being constant.
From the analysis here, it can be concluded that non-Gaussianity can
be large when the curvaton also contributes to the primordial
fluctuations without spoiling successful values of  $n_s$ and $r$.

Next we discuss the case with $n=4$. In fact, this model is on the verge
of the exclusion even by the WMAP3 data alone.  
If we assume $N_{\rm inf}=60$, 
the inflationary parameters become $(n_s, r) = (0.95,0.27)$,
which are marginal at the current data. 
However, by introducing the
curvaton in the model, the spectral index is modified to be shifted to
the scale-invariant one and the tensor-to-scalar ratio is
suppressed. Thus it can liberate the model.  In
Fig.~\ref{fig:chaotic4_ns}, contours of $n_s$ and $r$ are
shown in the $s$--$\sigma_\ast$ plane. Here, in our setting,
$N_{\rm inf} = 60.3$, which yields $(n_s, r) = (0.950,0.265)$ without
the curvaton contribution.  When the effects of the curvaton are small,
the values of $n_S$ and $r$ fall outside the allowed region. 
However, in particular, around the region where
$\sigma_\ast \ll M_{\rm pl}$ and $p \ll1$, the effects of the curvaton are
significant to make the inflationary parameters preferable values in some 
parameter space. 
To show in what cases the inflation model is liberated by the curvaton,
the excluded region from the WMAP3 data
is represented by the shaded region in Fig.~\ref{fig:chaotic4_fnl}.
Namely, the region without the shade is 
allowed by the data.  In the same figure, we also plot contours of  the non-linearity
parameters $f_{\rm NL}, \tau_{\rm NL}$ and $g_{\rm NL}$. 
Interestingly,
there is a parameter space where the model can avoid the exclusion by
the data while generating large non-Gaussianity.

As for the case with $n=6$, this model is completely ruled out by
current observations without the curvaton. When $N_{\rm inf}=60$ and $50$, the
spectral index and the tensor-to-scalar ratio are 
$(n_s, r) =(0.93,0.4)$ and $(0.92,0.48)$. 
However, 
in the same way as the case with $n=4$,
this model can be made to be allowed with the help of the
curvaton. In Fig.~\ref{fig:chaotic6_ns}, contours of $n_s$
and $r$ are shown. In our setting, $N_{\rm inf} = 64.0$,
which yields $(n_s, r) = (0.937,0.375)$ without the curvaton
contribution. In Fig.~\ref{fig:chaotic6_fnl}, contours of the non-linearity
parameters are shown along with the region excluded by the data in the
$s$--$\sigma_\ast$ plane. Since the original model of inflation is
strongly disfavored by the data, the allowed region in
Fig.~\ref{fig:chaotic6_fnl} is smaller compared to the counterpart for the
case with $n=4$. However, there is a region where the model is relaxed
to be allowed by observations and large non-Gaussianity can be
generated.

%%%%%%%%%%%%%%%%%%%%%
\subsection{New inflation}

\begin{figure}[htb]
\begin{center}
\scalebox{1}{
\includegraphics{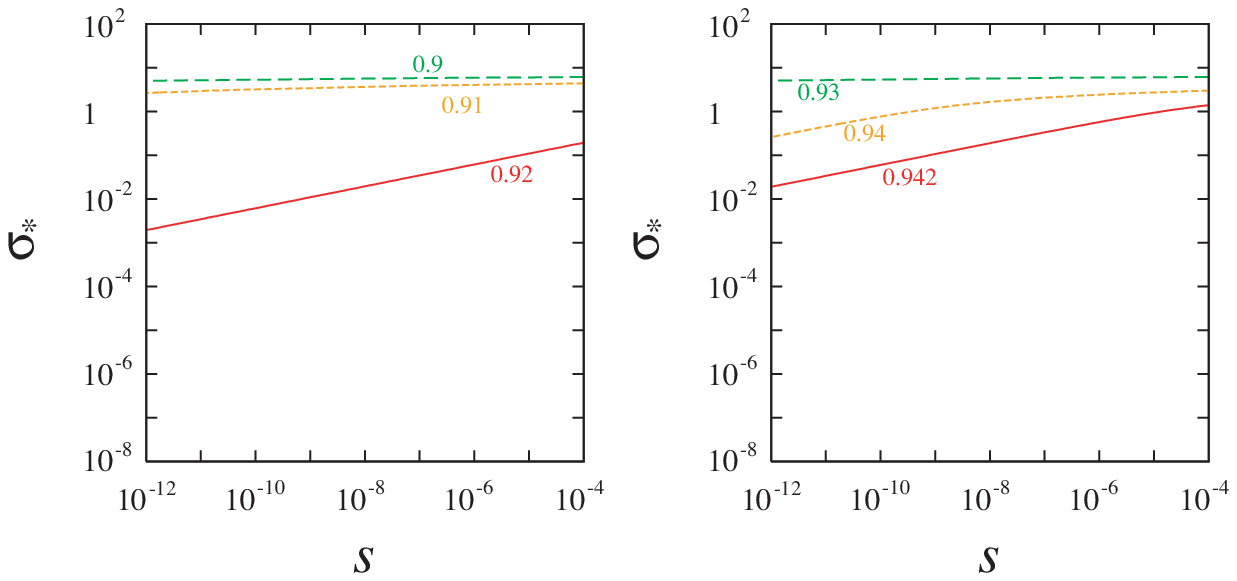}
}
\caption{Contours of $n_s$ in the $s$--$\sigma_\ast$ plane 
for the new inflation model with $n=3$ (left) and $n=4$ (right). 
We take $v=10^{-2}M_{\rm pl}$.
In the absence of the curvaton contribution, 
$n_s=0.920$ and $0.944$ for the case with $n=3, N_{\rm inf}=50.3$ 
and $n=4, N_{\rm inf}=53.3$ respectively.}
\label{fig:new_ns}
\end{center}
\end{figure}

Among various possibilities of the potential for the new inflation
model, we adopt the following form, which is motivated by the
supersymmetric models with the discrete $R$ symmetry of the
$Z_{\rm n}$ group \cite{Kumekawa:1994gx,Izawa:1996dv}, for definiteness:
\begin{equation}
\label{eq:V_new}
V = \lambda^2 v^4 
\left[ 
1 - 2 \left( \frac{\phi}{v} \right)^n + \left( \frac{\phi}{v} \right)^{2n} 
\right].
\end{equation}
During inflation, the last term in the RHS is irrelevant, thus
neglecting this term, we can write down the slow-roll parameters for the
potential above as
\begin{eqnarray}
\epsilon 
& = &  
2 n^2 \left( \frac{\phi_\ast}{v} \right)^{2(n-1)} \left( \frac{M_{\rm pl}}{v} \right)^2,
\\
\eta 
& = & 
- 2 n (n-1)
 \left( \frac{\phi_\ast}{v} \right)^{n-2} \left( \frac{M_{\rm pl}}{v} \right)^2,
\\
\xi^2 
&=&
4 n^2 (n-1)(n-2)
 \left( \frac{\phi_\ast}{v} \right)^{2(n-2)} \left( \frac{M_{\rm pl}}{v} \right)^4.
\end{eqnarray}
The number of $e$-folding is given approximately as
\begin{equation}
%\label{ }
N_{\rm inf} = \frac{1}{2n(n-2)} 
\left( \frac{v}{M_{\rm pl}} \right)^2 
 \left( \frac{\phi_\ast}{v} \right)^{-n+2}. 
\end{equation}
Since $\phi_\ast / v \ll 1$ during inflation, the slow-roll parameter
$\epsilon$ is much smaller than $\eta$ in magnitude,
which leads to the spectral index $n_s -1 \simeq 2\eta$.  
By using $N_{\rm inf}$, this can be written as
\begin{equation}
%\label{ }
n_s -1 \simeq -2 \frac{n-1}{n-2} \frac{1}{N_{\rm inf}}.
\end{equation}
Furthermore, the running of the spectral index 
$n_{\rm run} \simeq -2 \xi^2$ becomes
\begin{equation}
%\label{ }
n_{\rm run} \simeq -2 \frac{n-1}{n-2} \frac{1}{N_{\rm inf}^2}.
\end{equation}
The tensor-to-scalar ratio $r = 16 \epsilon$ is very small since
$\epsilon$ is suppressed as
\begin{equation}
%\label{ }
\epsilon = 2n^2  \left( \frac{M_{\rm pl}}{v} \right)^2 
\left[ \frac{1}{2n (n-2) N_{\rm inf} } \left( \frac{v}{M_{\rm pl}} \right) \right]^{2(n-1)/(n-2)}.
\end{equation}
If we take $n=3$ and $4$, the $\epsilon$ parameters are given by
$\epsilon = (v/M_{\rm pl})^6 / (72 N_{\rm inf}^4)$ and 
$\epsilon =4(v/M_{\rm pl})^4 / (27 N_{\rm inf}^3)$, respectively. 
Thus even if we take $v \sim M_{\rm pl}$, $\epsilon$ 
is suppressed with some powers of $N_{\rm inf}$ which is usually $50$--$60$.  
Hence both the tensor-to-scalar ratio and the running are very
small in this model.
The spectral indices for the cases with $n=3$ and $4$
are $n_s = 0.93$ and $0.95$ when $N_{\rm inf}=60$.
Thus this model is almost consistent with the current constraint 
from WMAP3 in terms of the spectral index and tensor mode.

One may speculate that,
adding a contribution from the curvaton,
non-Gaussianity becomes significantly large like in the case of 
the chaotic inflation.
However,
this does not happen in the new inflation.
This is because the slow-roll parameter $\epsilon$ is very small 
in the new inflation models.
Largest possible values for the non-linear
parameters which are given in Eqs.~\eqref{eq:fnl_max}, \eqref{eq:tnl_max} and
\eqref{eq:gnl_max} indicate that these values are proportional to
$\epsilon$.  
Thus inflation models with small $\epsilon$ cannot generate
large non-Gaussianity even by adding the curvaton.  

Furthermore, when $\epsilon$ is very small, 
the effects of the curvaton on $n_s$ and $r$ are also small. 
As seen from Eqs.~\eqref{eq:n_s}, \eqref{eq:n_run} and \eqref{eq:r}, the
effect of the curvaton on these quantities always appears in the
combination of $\epsilon Q_\sigma^2$. Thus even when $Q_\sigma$ is
large, this combination is small if $\epsilon$ is negligibly small.
To see this fact, 
contours of $n_s$ for the case with $n=3$ and $4$ are 
shown in Fig.~\ref{fig:new_ns}. 
In both case,
we take $v=10^{-2}M_{\rm pl}$.
Without the curvaton contribution, 
the $e$-folding numbers in our setting for $n=3$ and $4$ are given by
$N_{\rm inf} = 50.3$ and $53.3$ respectively. 
Then the spectral indices are $n_s = 0.920$ and $0.944$ respectively. 
As seen from the figure, 
the value of $n_s$ does not change much over the whole parameter space 
in the figure. 
This indicates that the effect of the curvaton cannot be large 
for a broad range of the curvaton parameters. 
In fact, when the initial amplitude for the curvaton is large, 
$n_s$ is modified a little because of the change of the number of $e$-folding
due to the second inflation. 
However, in any case, the figure clearly shows that 
the effect of the fluctuations of the curvaton is small for this kind of models.

\subsection{Hybrid inflation}

\begin{figure}[htb]
\begin{center}
\scalebox{1}{
\includegraphics{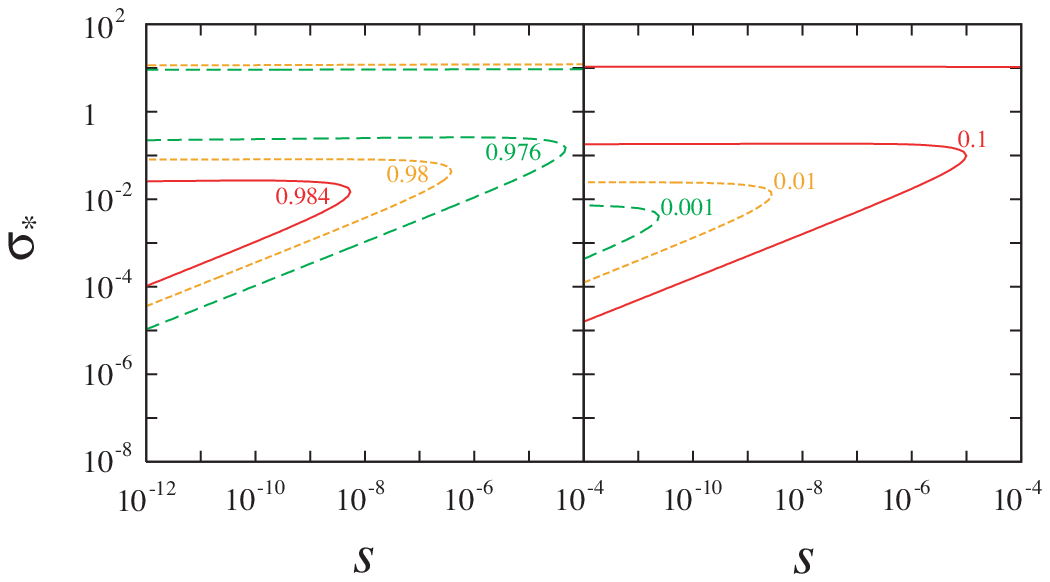} 
}
\caption{Contours of $n_s$ (left panel) and $r$ (right panel) 
    in the $s$--$\sigma_\ast$ plane for the
  hybrid inflation model with a quadratic potential. 
  In the absence of the curvaton contribution, the spectral index and tensor-to-scalar
  ratio become as $n_s=0.975$ and $r=0.118$ for $N_{\rm inf}=55.3$.}
\label{fig:hybrid2_ns}
\end{center}
\end{figure}

\begin{figure}[h]
\scalebox{1}{
\includegraphics{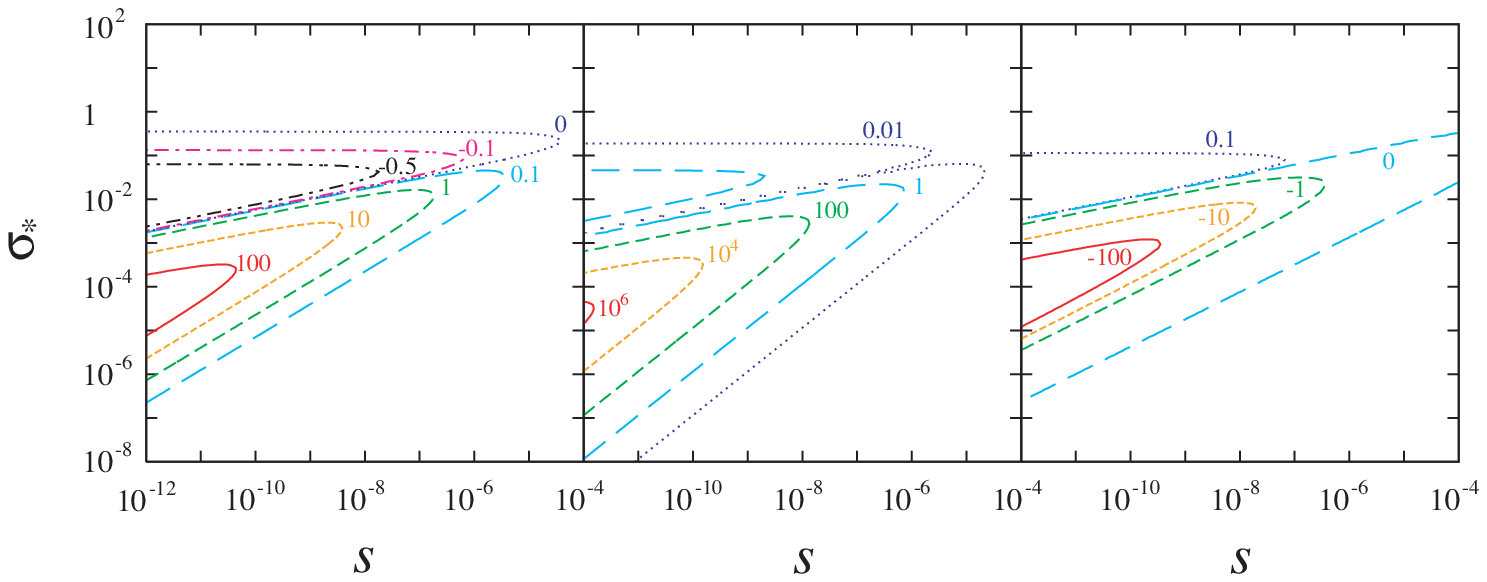} 
}
\caption{Contours of $f_{\rm NL}$ (left panel), $\tau_{\rm NL}$ (center
  panel) and $g_{\rm NL}$ (right panel) in the $s$--$\sigma_\ast$ plane
  for the hybrid inflation model with a quadratic potential. }
\label{fig:hybrid2_fnl}
\end{figure}

Hybrid inflation was originally proposed by Linde
\cite{Linde:1991km,Linde:1993cn} and many types of the potential have
been discussed in association with the gauge symmetry breaking
\cite{Copeland:1994vg,Dvali:1994ms,Lazarides:1996dv}. Here, we take the
original types of the potential for simplicity. The implications of
another field on other types of the potential like those in the
supersymmetric grand unified theory are discussed in the context of
modular inflation \cite{Lazarides:2007dg}. 

We consider the following quadratic type of potential,
\begin{equation}
\label{eq:V_hybrid2}
V=\alpha \left[ {(v^2-\sigma^2)}^2+\frac{m^2}{2}\phi^2
  +\frac{g^2}{2}\phi^2\sigma^2 \right].
\end{equation}
Here we set the parameters to be 
$v=10^{-2} M_{\rm pl}$, $m=2 \times 10^{-5} M_{\rm pl}$, 
and $g=2 \times 10^{-3}$. 
The parameter $\alpha$ is chosen to fit the WMAP normalization. 
For $\phi > \phi_{\rm end} = 2v/g=10 M_{\rm pl}$, 
the trajectory with $\sigma=0$ is stable and inflation
takes place along this trajectory with the effective potential given by
$V_{\rm eff}(\phi) = \alpha (v^4 + m^2 \phi^2/2)$. Then, the
slow-roll parameters are given by
\begin{equation}
\label{eq:slowparahyb2}
\epsilon = \frac{1}{2}\left( \frac{m^2 M_{\rm pl} \phi}
                                  {v^4+\frac12 m^2 \phi^2} \right)^2, ~~~
\eta = \frac{m^2 M_{\rm pl}^2}{v^4+\frac12 m^2 \phi^2}, ~~~ 
\xi^2 = 0.
\end{equation}
For the parameters above, without the curvaton contribution,
the number of $e$-folding, the spectral index and the tensor-to-scalar 
ratio are given by $N_{\rm inf} =55.3, n_s = 0.975$ and $r=0.118$. 
The running of the spectral index is negligible in this case.
Although these values are consistent with the WMAP3 data, 
to see how the curvaton affects these predictions, 
contours of $n_s$ and $r$ are shown in Fig.~\ref{fig:hybrid2_ns}.
Furthermore, contours of non-linearity parameters are also shown in
Fig.~\ref{fig:hybrid2_fnl}. 
In this case,
large non-Gaussianity can also be generated.

In the above choice of the parameters,
we have red-tilted power spectrum in the absence of the curvaton.
We can choose another set of parameters such that $m \ll gv$.
In this case,
we have $\eta \gg \epsilon$ and the spectrum becomes blue,
i.e.,
$n_s \approx 1+6\eta$,
in the absence of the curvaton.
If the curvaton contribution is small enough to satisfy $\epsilon M_{\rm pl}^2 Q_\sigma^2 \ll 1$,
then adding the curvaton yields no effects on the parameters 
as in the case of the new inflation.
On the other hand,
if the curvaton contribution is large enough to satisfy $\epsilon M_{\rm pl}^2 Q_\sigma^2 \gg 1$,
then the spectral index is given by
$n_s-1\approx -2\epsilon+\eta/(\epsilon M_{\rm pl}^2 Q_\sigma^2)$.
Because $\epsilon$ is very small ($\epsilon \ll \eta$) and
$\eta$ is positive in this model, 
the spectrum is almost scale invariant or blue-tilted,
and $r$ is negligibly small.
Hence,
for $m \ll gv$
in which case the spectrum is blue with negligibly small scalar-to-tensor ratio,
the model with such parameters is hardly allowed by WMAP3 (see Fig.~\ref{fig:wmap3})
even if the curvaton contributes to the primordial fluctuations.

%%%%%%%%%%%%%%%%%%%%%%%%%%%%%%%
\section{Summary and Discussions} \label{sec:summary}
%%%%%%%%%%%%%%%%%%%%%%%%%%%%%%%

In this paper, 
we studied primordial fluctuations in models where both the inflaton 
and the curvaton can contribute to the cosmic density fluctuations. 
By using the $\delta N$ formalism,
we provided a systematic formulation to discuss the primordial
curvature fluctuations and their non-Gaussianity in this kind of scenario. 
We gave the expressions for the spectral index $n_s$, its
running $n_{\rm run}$, the tensor-to-scalar ratio $r$ and the non-linearity
parameters $f_{\rm NL}, \tau_{\rm NL}$ and $g_{\rm NL}$ for the mixed
model.  In general, the values of these parameters can be obtained by
numerically calculating some set of equations, however, in some cases,
they are given analytically. 
We presented such analytic formulae in some limiting cases.
We found that very large non-Gaussianity ($f_{\rm NL} \gg 1$)
can be generated even if the contribution of the curvaton fluctuations 
to the total ones is comparable to or smaller than that of the inflaton fluctuations.
If the curvaton fluctuations are minor components,
then $\tau_{\rm NL}$ is enhanced by the  factor of 
$\zeta_{\rm inf}^2/\zeta_{\rm cur}^2 \gg 1$,
compared to $f_{\rm NL}^2$,
where $\zeta_{\rm inf}$ and $\zeta_{\rm cur}$ are the primordial
fluctuations from the inflaton and the curvaton respectively.
In such a case,
large non-Gaussian feature appears through the trispectrum 
rather than through the bispectrum.
We also derived the consistency relation between the bispectrum and the trispectrum
which holds in the limit of the large non-Gaussianity.
If the measurement of the trispectrum becomes available in the future,
this relation will provide a useful tool to distinguish the
mixed model of the inflaton and the curvaton from other scenarios 
which also generate large non-Gaussianity.

After we worked out the formulation for the calculation of inflationary
parameters including the non-linearity parameters, 
we investigated the issue for several concrete models of inflation: chaotic
inflation, new inflation and hybrid inflation.  By calculating the
spectral index and the tensor-to-scalar ratio, we can compare those with
cosmological observations.  
We showed that some inflation models already
excluded by the current data such as chaotic inflation with higher
polynomials $n=4$ and $6$ can be liberated depending on the parameters
which characterize the curvaton.  By presenting contours of $n_s$ and
$r$ in the $s$--$\sigma_\ast$ plane, we have explicitly shown in what
cases such models can be relaxed to be allowed by the data. 
We have also presented contours of the non-linearity parameters in the 
$s$--$\sigma_\ast$ plane. 
In particular,
it should be noticed that, the non-Gaussianity can be large due to the
contribution from the curvaton without conflicting with the data.

On the other hand, 
some inflation models such as the new inflation 
where the slow-roll parameter $\epsilon$ is very
small are scarcely affected by the curvaton.
The effects of the curvaton on $n_s$ and $r$ appear
as in the combination of $\epsilon Q_\sigma^2$, thus the size of
$\epsilon$ is crucial when one considers how the curvaton affects the
inflationary predictions. 
Furthermore, non-Gaussianity cannot be large in this model 
even if we introduce the curvaton because the $\epsilon$ parameter
is very small in this model. 
As shown in Eqs.~(\ref{eq:f_NL})$\sim$(\ref{eq:g_NL}),
the curvaton contribution to the non-linearity parameters is always
associated with $\epsilon$.
Thus inflation models with very small $\epsilon$ cannot generate 
large non-Gaussianity even with the curvaton.

Although we have concentrated on adiabatic fluctuations thus far, 
isocurvature fluctuations can be produced  in
principle, depending on the thermal history of the universe. 
It can happen when dark matter or baryon was generated (and its
number is conserved) before the decay of the curvaton or from the decay
products of the curvaton \cite{Moroi:2002rd,Lyth:2002my,Lyth:2003ip}. In the case
where dark matter and baryon were produced after the decay of the
curvaton, no isocurvature perturbation is generated even in our mixed
model. In this paper, we have considered this kind of case to avoid 
complexities coming from the contamination of isocurvature modes
when we interpret observational signatures.
However, here we comment on the cases where some isocurvature 
fluctuations are generated.
First of all, 
since pure cold dark matter (CDM)/bayron isocurvature modes (regardless of its correlation with 
adiabatic mode) are severely constrained by 
observations, the pure curvaton scenario with isocurvature fluctuations 
such as the above mentioned cases
are disfavored from current observations.
However, it should be noted here that 
the constraint can be relaxed when adiabatic fluctuations
from the inflaton contribute to the total density fluctuations, 
which is the case for the mixed scenario.
This is simply because the contribution from adiabatic mode
decreases the fraction of the isocurvature 
contribution to the total  fluctuations.
Thus the mixed scenario may have interesting implications
on scenarios with isocurvature fluctuations.  For
example, even in the case where dark matter was generated before the decay of
the curvaton, the mixed scenario is still allowed if adiabatic
fluctuations from the inflaton is bigger than that from the curvaton to some extent.
Similarly, although a scenario where dark matter was
generated from the decay products of the curvaton is 
severely constrained by the argument of isocurvature fluctuations,
as in the case discussed above, the constraint
is significantly relaxed in the mixed scenario if adiabatic fluctuation
from the inflaton is much bigger than that from the curvaton. 
It should be noticed that large non-Gaussianity would not be produced 
when the contribution from the inflaton is large compared to that from 
the curvaton.
On the other hand, when the contribution from the curvaton 
with isocurvature mode is large, such a scenario is severely constrained 
by the consideration of isocurvature mode, which corresponds to 
the case where the fraction of energy density of the curvaton to the 
total one becomes comparable to that of the inflaton. 
Remind that  large 
non-Gaussianity can be generated when the fraction of energy 
density of the curvaton to the total one is very small.
Thus non-Gaussianity would not be so large in this case too.

Since a simple single-field inflation model can only generate very small
non-Gaussianity, if the evidence of non-Gaussianity, which was
recently reported in Ref.~\cite{Yadav:2007yy}, is established in
the future, this mechanism of generating large non-Gaussianity will be
very interesting. Although we concentrate on the mixed model of
the inflaton and the curvaton in this paper, other mixed model may also be
worth investigating. 
In particular, large non-Gaussianity can also be generated in the modulated reheating
\cite{Zaldarriaga:2003my,Bartolo:2003bz,Vernizzi:2003vs,Suyama:2007bg}
and preheating scenarios \cite{Jokinen:2005by,Barnaby:2006km,Chambers:2007se} so that the
implications of such a mixed model on the inflationary parameters and
non-Gaussianity are of great interest and will be presented elsewhere \cite{Ichikawa:2008ne}.

\bigskip
\bigskip

\noindent {\bf Acknowledgments:} 
This work is supported in part by the Sumitomo Foundation (T.T.) and the
Grant-in-Aid for Scientific Research from the Ministry of Education,
Science, Sports, and Culture of Japan No.\,18840010 (K.I.),
No.\,19740145 (T.T.), No.\,18740157, and No.\,19340054 (M.Y.).
TS thanks to the computer system at the Yukawa Institute for Theoretical Physics,
Kyoto University, 
for the numerical calculations.

\pagebreak

\appendix 

\noindent
{\bf \Large Appendix}
\section{Value of $\sigma$ at the onset of the curvaton oscillations}\label{sec:alpha}

In the main text, we defined $g(\sigma_\ast)$ with the energy density of the
curvaton at the onset of the oscillations as
\begin{equation}
%\label{ }
\rho_\sigma ( N_m)  =\frac{1}{2} m_\sigma^2 g (\sigma_\ast)^2, 
\end{equation}
with $g(\sigma_\ast) = \alpha \sigma_\ast$. We assume that the universe
is radiation-dominated when the curvaton begins to oscillate, thus the
equation of motion for $\sigma$ can be written as
\begin{equation}
%\label{ }
\ddot{\sigma} + \frac{3}{2t} \dot{\sigma} + m_\sigma^2 \sigma =0,
\end{equation}
where we used $H = 1/(2t) $ during radiation-dominated epoch.  The
solution for this equation with the initial value $\sigma (t_m) =
\sigma_\ast$ is given by
\begin{equation}
%\label{ }
\sigma (t)  = \frac{2^{1/4} \Gamma (5/4) \sigma_\ast}{(mt)^{1/4}} J_{1/4} (mt),
\end{equation}
where $J$ is the Bessel function. Using the asymptotic form of
$J_{1/4}$, the energy density $\rho_\sigma$ for $mt \gg 1$ becomes 
\begin{equation}
%\label{ }
\rho_\sigma (t) \simeq \frac{\sqrt{2} m_\sigma^2 \sigma_\ast^2}{\pi (mt)^{3/2}}
\Gamma (5/4).
\end{equation}
Thus $\alpha$ can be given by 
\begin{equation}
%\label{ }
\alpha = \frac{2\sqrt{2}}{\sqrt{\pi}} \Gamma (5/4) \simeq 1.45.
\end{equation}

%%%%%%%%%%%%%%%%%%%%%%%%%%%%%%%%%%%%%%%%%%%%
\section{Analytic expression of $F(p)$ for $p \ll 1$}\label{sec:F_p}
%%%%%%%%%%%%%%%%%%%%%%%%%%%%%%%%%%%%%%%%%%%%

In this Appendix, we give a detailed description how we obtain an 
analytic expression of $F(p)$ for the case with $p \ll 1$. We also assume that 
the value of $s=\Gamma_\sigma / m_\sigma$ is also very small. 
$F(p)$ is given as
\begin{eqnarray}
F(p)=\int_0^\infty dN~e^{4N} \frac{\Gamma_\sigma}{H(N)} 
\frac{\rho_\sigma (N)}{\rho_{r0}}.
\end{eqnarray}
To evaluate $F(p)$, we have to follow the evolution of energy densities of the curvaton
and radiation after the curvaton begins to oscillate for the case we are going to consider
here. Thus we take the initial time $N=0$ when $H$ is equal to $m_\sigma$
and the curvaton field can be considered to be dust. In this case, 
the background equations we need to follow are:
\begin{eqnarray}
&&
\frac{d}{dN} \rho_r+4\rho_r
=\frac{\Gamma_\sigma}{H}\rho_\sigma, \label{eqr}\\
&&
\frac{d}{dN}\rho_\sigma+3 \rho_\sigma
=
-\frac{\Gamma_\sigma}{H}\rho_\sigma, \label{eqs}\\
&&
H^2=\frac{1}{3M_{\rm pl}^2} (\rho_r+\rho_\sigma), \label{eqH}
\end{eqnarray}
where we use the number of $e$-folding as a time variable.
We solve this set of equations with the initial conditions:
\begin{eqnarray}
\rho_r(0)=\rho_{r0},~~~~~\rho_\sigma (0)=\rho_{\sigma 0}=\frac{\alpha^2 m_\sigma^2 \sigma_*^2}{2},~~~~~H(0)=H_0=m_\sigma.
\end{eqnarray}

We assume that curvaton energy density is negligibly
small throughout the evolution.
Hence we treat $\rho_{\sigma 0}$ as the expansion parameter of
order $p$,
where $p$ is infinitesimal.
Then we can expand $\rho_r,~\rho_\sigma, H$ with respect to $p$ as
\begin{eqnarray}
&&\rho_r=\rho_r^{(0)}+\rho_r^{(1)}+\rho_r^{(2)}+\cdots, \\
&&\rho_\sigma=\rho_\sigma^{(0)}+\rho_\sigma^{(1)}+\rho_\sigma^{(2)}+\cdots, \\
&&H=H^{(0)}+H^{(1)}+H^{(2)}+\cdots,
\end{eqnarray}
where $\rho_r^{(n)},~\rho_\sigma^{(n)},~H^{(n)} ={\cal O}(p^n)$.
We also expand $F(p)$ with respect to $p$ as
\begin{eqnarray}
F=F^{(0)}+F^{(1)}+F^{(2)}+\cdots.
\end{eqnarray}
Obviously, $F^{(0)}$ vanishes. 
We are going to obtain the expression for $F(p)$ by solving 
the above set of equations order by order in the following. \\

\noindent
{\bf $\bullet$ $0$th order in $p$}\\
Since the energy density of the curvaton field 
is  considered to be the order of ${\cal O}(p)$,
at the $0$-th order, it vanishes as $\rho_\sigma^{(0)}=0$.
Other quantities can be  easily obtained.
\begin{eqnarray}
\rho_r^{(0)}=\rho_{r0}e^{-4N},~~~~~H^{(0)}=H_0 e^{-2N}.
\end{eqnarray}

\noindent
{\bf $\bullet$ 1st  order in $p$}\\
From Eq.~(\ref{eqs}),
the equation for $\rho_\sigma^{(1)}$ can be written as
\begin{eqnarray}
\frac{d}{dN} \rho_\sigma^{(1)}+3 \rho_\sigma^{(1)}=-s e^{2N} \rho_\sigma^{(1)}.
\end{eqnarray}
By solving this equation, the energy density of the curvaton in the 1st order 
is given by
\begin{eqnarray}
\rho_\sigma^{(1)}=\rho_{\sigma 0} \exp \bigg[ -3N-\frac{s}{2} (e^{2N}-1) \bigg]. \label{sols1}
\end{eqnarray}
At the 1st order in $p$, $F(p)$ can be given by
\begin{equation}
F^{(1)}(p) = \int_0^\infty dN e^{4N} \frac{\Gamma_\sigma}{H^{(0)}} 
\frac{\rho_\sigma^{(1)} }{\rho_{r0} }.
\end{equation}
By substituting the expressions for $\rho_\sigma^{(1)}$ and $H^{(0)}$ 
into the above equation,  we obtain
\begin{equation}
F^{(1)}(p) = 
\int_0^\infty dN e^{4N} \frac{\Gamma_\sigma}{H_0 e^{-2N}} 
\frac{\rho_{\sigma 0} }{3H_0^2 M_{\rm pl}^2} 
\exp \left[ 
 -3 N - \frac{\Gamma_\sigma}{2 H_0} \left( e^{2N} - 1 \right)
\right].
\end{equation}
The integral which appears in this equation 
can be done analytically as,
\begin{equation}
\int_0^\infty dN 
\exp \left[ 
 3 N - \frac{\Gamma_\sigma}{2 m_\sigma} \left( e^{2N} - 1 \right)
\right]
=
\sqrt{\frac{1}{2 s^3}}
\left( \sqrt{2s} + \sqrt{\pi} e^{s/2} \left( 1 - {\rm Erf} (\sqrt{s/2} ) \right) \right).
\end{equation}
where ${\rm Erf}(x)$ is the error function and defined by 
\begin{equation}
%\label{ }
{\rm Erf} (x) \equiv \frac{2}{\sqrt{\pi}} \int_0^x dt e^{-t^2}.
\end{equation}
By expanding the error function with $s$ and taking the leading order, 
we can obtain $F$ at the first order in $p$ as
\begin{equation}
%\label{ }
F^{(1)}(p) =
 \frac{1}{6} \sqrt{\frac{\pi}{2}} \alpha^2 p. 
\end{equation}

To evaluate the function $F(p)$ at the second order, we
need the expressions for $\rho_r^{(1)}$ and $H^{(1)}$. 
For the energy density of radiation for the 1st order, from Eq.~(\ref{eqr}),
we have the equation for $\rho_r^{(1)}$ as
\begin{eqnarray}
\frac{d}{dN} \rho_r^{(1)}+4 \rho_r^{(1)}=s e^{2N} \rho_\sigma^{(1)}.
\end{eqnarray}
Using Eq.~(\ref{sols1}), the solution of this equation can be formally written as
\begin{eqnarray}
\rho_r^{(1)}=s \rho_{\sigma 0} e^{-4N} \int_0^N dN'~\exp \left( 3N'-\frac{s}{2} (e^{2N'}-1) \right). \label{solr1}
\end{eqnarray}
By plugging this expression into the Friedmann equation,
the Hubble parameter can be given by, in the 1st order,
\begin{eqnarray}
\frac{H^{(1)}}{H^{(0)}}=\frac{1}{2} \frac{\rho_{\sigma 0}}{\rho_{r0}} e^{\frac{s}{2}} \Bigg[ s \int_0^N dN'~\exp \left( 3N'-\frac{s}{2}e^{2N'} \right)+\exp \left( N-\frac{s}{2}e^{2N} \right) \Bigg]. \label{solH1}
\end{eqnarray}

\noindent
{\bf $\bullet$ 2nd order in $p$}\\
Now we calculate the function $F(p)$ for the second order in $p$.
$F^{(2)}(p)$ can be written as
\begin{eqnarray}
F^{(2)}=s \int_0^\infty dN~e^{6N} \frac{\rho_\sigma^{(2)}}{\rho_{r0}}-s \int_0^\infty dN ~e^{6N} \frac{H^{(1)}}{H^{(0)}} \frac{\rho_\sigma^{(1)}}{\rho_{r0}}. \label{F2}
\end{eqnarray}
To evaluate the RHS, 
we do not need $\rho_r^{(2)}$ and $H^{(2)}$ because
these do not appear in $F(p)$ up to second order in $p$, but 
we need to have the expression for $\rho_\sigma^{(2)}$.
From Eq.~(\ref{eqs}), we have an equation for $\rho_\sigma^{(2)}$ as
\begin{eqnarray}
\frac{d}{dN} \rho_\sigma^{(2)}+3 \rho_\sigma^{(2)}=-s e^{2N} \rho_\sigma^{(2)}+se^{2N} \frac{H^{(1)}}{H^{(0)}} \rho_\sigma^{(1)}.
\end{eqnarray}
Using the first order solutions obtained above,
$\rho_\sigma^{(2)}$ can be written as
\begin{eqnarray}
&&\rho_\sigma^{(2)}=\frac{s}{2} \frac{\rho_{\sigma 0}}{\rho_{r0}} \rho_{\sigma 0}e^s \exp \left( -3N-\frac{s}{2} e^{2N} \right) \int_0^N dN'~e^{2N'} \nonumber \\
&&\hspace{10mm} \times \Bigg[ s \int_0^{N'}dN''~\exp \left( 3N''-\frac{s}{2} e^{2N''} \right)+\exp \left( N'-\frac{s}{2} e^{2N'} \right) \Bigg]. 
\label{eq:rho_r2}
\end{eqnarray}
In principle, $F^{(2)}$ can be obtained by plugging Eqs.~\eqref{solr1}, \eqref{solH1}
and \eqref{eq:rho_r2} into Eq.~\eqref{F2}, 
the expression becomes lengthy and complicated, we consider each 
term in the RHS of Eq.~\eqref{F2} in each.

We denote the first term in Eq.~(\ref{F2}) as $T_1$.
Using the formulae for $\rho_r^{(1)}$ and $H^{(1)}$, 
$T_1$ can be written as
\begin{eqnarray}
&&T_1=\frac{s^2}{2} {\left( \frac{\rho_{\sigma 0}}{\rho_{r0}} \right)}^2 e^s \int_0^\infty dN~\exp \left( 3N-\frac{s}{2} e^{2N} \right) \int_0^N dN'~e^{2N'} \nonumber \\
&&\hspace{10mm}\times \Bigg[ s\int_0^{N'} dN''~\exp \left( 3N''-\frac{s}{2} e^{2N''} \right)+\exp \left( N'-\frac{s}{2} e^{2N'} \right) \Bigg]. \label{T1}
\end{eqnarray}
Although we need to perform a triple integral,
we can make use the following formula\footnote{
This equation can be verified by changing the variable as
\begin{eqnarray}
\sqrt{ \frac{s}{2} }e^N =y.
\end{eqnarray}
}:
\begin{eqnarray}
\int_0^N dN'~\exp \left( 3N'-\frac{s}{2} e^{2N'} \right) 
\simeq
{\left( \frac{2}{s} \right)}^{3/2} \Bigg[ -\frac{1}{2} \sqrt{ \frac{s}{2} } \exp \left( N-\frac{s}{2} e^{2N} \right)+\frac{\sqrt{\pi}}{4} {\rm Erf} \left( \sqrt{ \frac{s}{2} } e^N \right) \Bigg], \notag \\
\end{eqnarray}
where we have neglected higher order terms in $s$ because
we are interested in the case $s \ll 1$.
Using this formula,
the first term in Eq.~(\ref{T1}) can be written as $\beta_1/s^3$
(omitting the factor appearing outside the integral and taking only
leading order term in $s$)
where $\beta_1$ is given by
\begin{eqnarray}
\beta_1 \equiv 8 \int_0^\infty dy~y^2 e^{-y^2} \int_0^y dx \left(-x^2 e^{-x^2}+ \frac{ \sqrt{\pi}}{2} x {\rm Erf} (x) \right).
\end{eqnarray}
After some calculations, we find $\beta_1=1$.
The second term in Eq.~(\ref{T1}) can be calculated as $\pi/(4s^3)$.
Hence $T_1$ is given by
\begin{eqnarray}
T_1=\frac{1}{2s} {\left( \frac{\rho_{\sigma 0}}{\rho_{r0}} \right)}^2 \left( 1+\frac{\pi}{4} \right).
\end{eqnarray}

Next we evaluate the second term in Eq.~(\ref{F2}) which is denoted as $T_2$.
By using the expressions for $\rho_\sigma^{(1)}$ and $H^{(1)}$, 
$T_2$ can be written as
\begin{eqnarray}
&&T_2=\frac{s}{2} {\left( \frac{\rho_{\sigma 0}}{\rho_{r0}} \right)}^2 e^s \int_0^\infty dN~\exp \left( 3N-\frac{s}{2} e^{2N} \right) \nonumber \\ 
&&\hspace{10mm} \times \Bigg[ s\int_0^N dN'~\exp \left( 3N'-\frac{s}{2} e^{2N'} \right)+\exp \left( N-\frac{s}{2} e^{2N} \right) \Bigg]. \label{T2}
\end{eqnarray}
Similar calculations to the case with $T_1$ yields
\begin{eqnarray}
T_2=\frac{1}{4s} \left( 1+\frac{\pi}{2} \right) {\left( \frac{\rho_{\sigma 0}}{\rho_{r0}} \right)}^2,
\end{eqnarray}
where we have again dropped higher order terms in $s$.
Thus we obtain the expression for $F^{(2)}$ as
\begin{eqnarray}
F^{(2)}=T_1-T_2=\frac{\alpha^4}{144} p^2.
\end{eqnarray}


\begin{thebibliography}{100}


%\cite{Spergel:2006hy}
\bibitem{Spergel:2006hy}
  D.~N.~Spergel {\it et al.}  [WMAP Collaboration],
  %``Wilkinson Microwave Anisotropy Probe (WMAP) three year results:
  %Implications for cosmology,''
  Astrophys.\ J.\ Suppl.\  {\bf 170}, 377 (2007)
  [arXiv:astro-ph/0603449].
  %%CITATION = APJSA,170,377;%%

%\cite{Tegmark:2006az}
\bibitem{Tegmark:2006az}
  M.~Tegmark {\it et al.}  [SDSS Collaboration],
  %``Cosmological Constraints from the SDSS Luminous Red Galaxies,''
  Phys.\ Rev.\  D {\bf 74}, 123507 (2006)
  [arXiv:astro-ph/0608632].
  %%CITATION = PHRVA,D74,123507;%%

%\cite{Yadav:2007yy}
\bibitem{Yadav:2007yy}
  A.~P.~S.~Yadav and B.~D.~Wandelt,
  %``Detection of primordial non-Gaussianity (fNL) in the WMAP 3-year data at
  %above 99.5% confidence,''
  arXiv:0712.1148 [astro-ph].
  %%CITATION = ARXIV:0712.1148;%%

%\bibitem{curvaton}
\bibitem{Enqvist:2001zp}
K.~Enqvist and M.~S.~Sloth,
%``Adiabatic CMB perturbations in pre big bang string cosmology,''
Nucl.\ Phys.\ B {\bf 626}, 395 (2002)
[arXiv:hep-ph/0109214];

\bibitem{Lyth:2001nq}
D.~H.~Lyth and D.~Wands,
%``Generating the curvature perturbation without an inflaton,''
Phys.\ Lett.\ B {\bf 524}, 5 (2002)
[arXiv:hep-ph/0110002];

\bibitem{Moroi:2001ct}
T.~Moroi and T.~Takahashi,
%``Effects of cosmological moduli fields on cosmic microwave background,''
Phys.\ Lett.\ B {\bf 522}, 215 (2001)
[Erratum-ibid.\ B {\bf 539}, 303 (2002)]
[arXiv:hep-ph/0110096].
%%%

%\cite{Dvali:2003em}
\bibitem{Dvali:2003em}
  G.~Dvali, A.~Gruzinov and M.~Zaldarriaga,
  %``A new mechanism for generating density perturbations from inflation,''
  Phys.\ Rev.\  D {\bf 69}, 023505 (2004)
  [arXiv:astro-ph/0303591].
  %%CITATION = PHRVA,D69,023505;%%
  
%\cite{Kofman:2003nx}
\bibitem{Kofman:2003nx}
  L.~Kofman,
  %``Probing string theory with modulated cosmological fluctuations,''
  arXiv:astro-ph/0303614.
  %%CITATION = ASTRO-PH/0303614;%%  

\bibitem{Dimopoulos:2002kt}
  K.~Dimopoulos and D.~H.~Lyth,
  %``Models of inflation liberated by the curvaton hypothesis,''
  Phys.\ Rev.\ D {\bf 69}, 123509 (2004)
  [arXiv:hep-ph/0209180].

%\cite{Endo:2003fr}
\bibitem{Endo:2003fr}
  M.~Endo, M.~Kawasaki and T.~Moroi,
  %``Cosmic string from D-term inflation and curvaton,''
  Phys.\ Lett.\  B {\bf 569}, 73 (2003)
  [arXiv:hep-ph/0304126].
  %%CITATION = PHLTA,B569,73;%%
  
%\cite{Lazarides:2004we}
\bibitem{Lazarides:2004we}
  G.~Lazarides, R.~R.~de Austri and R.~Trotta,
  %``Constraints on a mixed inflaton and curvaton scenario for the  generation
  %of the curvature perturbation,''
  Phys.\ Rev.\  D {\bf 70}, 123527 (2004)
  [arXiv:hep-ph/0409335].
  %%CITATION = PHRVA,D70,123527;%%      
    
%\cite{Dimopoulos:2004yb}
\bibitem{Dimopoulos:2004yb}
  K.~Dimopoulos, D.~H.~Lyth and Y.~Rodriguez,
  %``Low scale inflation and the curvaton mechanism,''
  JHEP {\bf 0502}, 055 (2005)
  [arXiv:hep-ph/0411119].
  %%CITATION = JHEPA,0502,055;%%

%\cite{Rodriguez:2004yc}
\bibitem{Rodriguez:2004yc}
  Y.~Rodriguez,
  %``Low scale inflation and the immediate heavy curvaton decay,''
  Mod.\ Phys.\ Lett.\  A {\bf 20}, 2057 (2005)
  [arXiv:hep-ph/0411120].
  %%CITATION = MPLAE,A20,2057;%%

\bibitem{Langlois:2004nn}
D.~Langlois and F.~Vernizzi,
%``Mixed inflaton and curvaton perturbations,''
Phys.\ Rev.\ D {\bf 70}, 063522 (2004)
[arXiv:astro-ph/0403258].

%\cite{Moroi:2005kz}
\bibitem{Moroi:2005kz}
  T.~Moroi, T.~Takahashi and Y.~Toyoda,
  %``Relaxing constraints on inflation models with curvaton,''
  Phys.\ Rev.\  D {\bf 72}, 023502 (2005)
  [arXiv:hep-ph/0501007].
  %%CITATION = PHRVA,D72,023502;%%
  
%\cite{Moroi:2005np}
\bibitem{Moroi:2005np}
  T.~Moroi and T.~Takahashi,
  %``Implications of the curvaton on inflationary cosmology,''
  Phys.\ Rev.\  D {\bf 72}, 023505 (2005)
  [arXiv:astro-ph/0505339].
  %%CITATION = PHRVA,D72,023505;%%

%\cite{Starobinsky:1986fxa}
\bibitem{Starobinsky:1986fxa}
  A.~A.~Starobinsky,
  %``Multicomponent de Sitter (Inflationary) Stages and the Generation of
  %Perturbations,''
  JETP Lett.\  {\bf 42} (1985) 152
  [Pisma Zh.\ Eksp.\ Teor.\ Fiz.\  {\bf 42} (1985) 124].
  %%CITATION = ZFPRA,42,124;%%

%\cite{Sasaki:1995aw}
\bibitem{Sasaki:1995aw}
  M.~Sasaki and E.~D.~Stewart,
  %``A General Analytic Formula For The Spectral Index Of The Density
  %Perturbations Produced During Inflation,''
  Prog.\ Theor.\ Phys.\  {\bf 95}, 71 (1996)
  [arXiv:astro-ph/9507001]. 
  %%CITATION = ASTRO-PH 9507001;%%

%\cite{Sasaki:1998ug}
\bibitem{Sasaki:1998ug}
  M.~Sasaki and T.~Tanaka,
  %``Super-horizon scale dynamics of multi-scalar inflation,''
  Prog.\ Theor.\ Phys.\  {\bf 99}, 763 (1998)
  [arXiv:gr-qc/9801017].
  %%CITATION = GR-QC 9801017;%%  

%\cite{Lyth:2004gb}
\bibitem{Lyth:2004gb}
  D.~H.~Lyth, K.~A.~Malik and M.~Sasaki,
  %``A general proof of the conservation of the curvature perturbation,''
  JCAP {\bf 0505}, 004 (2005)
  [arXiv:astro-ph/0411220].
  %%CITATION = JCAPA,0505,004;%%

%\cite{Seery:2005gb}
\bibitem{Seery:2005gb}
  D.~Seery and J.~E.~Lidsey,
  %``Primordial non-gaussianities from multiple-field inflation,''
  JCAP {\bf 0509}, 011 (2005)
  [arXiv:astro-ph/0506056].
  %%CITATION = JCAPA,0509,011;%%

%\cite{Seery:2006vu}
\bibitem{Seery:2006vu}
  D.~Seery, J.~E.~Lidsey and M.~S.~Sloth,
  %``The inflationary trispectrum,''
  JCAP {\bf 0701}, 027 (2007)
  [arXiv:astro-ph/0610210].
  %%CITATION = JCAPA,0701,027;%%

%\cite{Arroja:2008ga}
\bibitem{Arroja:2008ga}
  F.~Arroja and K.~Koyama,
  %``Non-gaussianity from the trispectrum in general single field inflation,''
  arXiv:0802.1167 [hep-th].
  %%CITATION = ARXIV:0802.1167;%%
  
\bibitem{Lyth:2005fi}
  D.~H.~Lyth and Y.~Rodriguez,
  %``The inflationary prediction for primordial non-gaussianity,''
  Phys.\ Rev.\ Lett.\  {\bf 95}, 121302 (2005)
  [arXiv:astro-ph/0504045].
  %%CITATION = ASTRO-PH 0504045;%%
   
%\cite{Alabidi:2005qi}
\bibitem{Alabidi:2005qi}
  L.~Alabidi and D.~H.~Lyth,
  %``Inflation models and observation,''
  JCAP {\bf 0605}, 016 (2006)
  [arXiv:astro-ph/0510441].
  %%CITATION = JCAPA,0605,016;%%

%\cite{Byrnes:2006vq}
\bibitem{Byrnes:2006vq}
  C.~T.~Byrnes, M.~Sasaki and D.~Wands,
  %``The primordial trispectrum from inflation,''
  Phys.\ Rev.\  D {\bf 74}, 123519 (2006)
  [arXiv:astro-ph/0611075].
  %%CITATION = PHRVA,D74,123519;%%

%\cite{Suyama:2007bg}
\bibitem{Suyama:2007bg}
  T.~Suyama and M.~Yamaguchi,
  %``Non-Gaussianity in the modulated reheating scenario,''
  Phys.\ Rev.\  D {\bf 77}, 023505 (2008)
  [arXiv:0709.2545 [astro-ph]].
  %%CITATION = PHRVA,D77,023505;%%

%\cite{Polarski:1992dq}
\bibitem{Polarski:1992dq}
  D.~Polarski and A.~A.~Starobinsky,
  %``Spectra of perturbations produced by double inflation with an intermediate
  %matter dominated stage,''
  Nucl.\ Phys.\  B {\bf 385}, 623 (1992).
  %%CITATION = NUPHA,B385,623;%%

  %\cite{Malik:2002jb}
\bibitem{Malik:2002jb}
  K.~A.~Malik, D.~Wands and C.~Ungarelli,
  %``Large-scale curvature and entropy perturbations for multiple interacting
  %fluids,''
  Phys.\ Rev.\  D {\bf 67}, 063516 (2003)
  [arXiv:astro-ph/0211602].
  %%CITATION = PHRVA,D67,063516;%%

 %\cite{Gupta:2003jc}
\bibitem{Gupta:2003jc}
  S.~Gupta, K.~A.~Malik and D.~Wands,
  %``Curvature and isocurvature perturbations in a three-fluid model of
  %curvaton decay,''
  Phys.\ Rev.\  D {\bf 69}, 063513 (2004)
  [arXiv:astro-ph/0311562].

%\cite{Huang:2008ze}
\bibitem{Huang:2008ze}
  Q.~G.~Huang,
  %``Large Non-Gaussianity Implication for Curvaton Scenario,''
  arXiv:0801.0467 [hep-th].
  %%CITATION = ARXIV:0801.0467;%%

\bibitem{Turner:1983he}
M.~S.~Turner,
%``Coherent Scalar Field Oscillations In An Expanding Universe,''
Phys.\ Rev.\ D {\bf 28}, 1243 (1983).
   
%\cite{Linde:1983gd}
\bibitem{Linde:1983gd}
  A.~D.~Linde,
  %``Chaotic Inflation,''
  Phys.\ Lett.\  B {\bf 129}, 177 (1983).
  %%CITATION = PHLTA,B129,177;%%

%\cite{Goncharov:1983mw}
\bibitem{Goncharov:1983mw}
  A.~B.~Goncharov and A.~D.~Linde,
  %``Chaotic Inflation In Supergravity,''
  Phys.\ Lett.\  B {\bf 139}, 27 (1984).
  %%CITATION = PHLTA,B139,27;%%

%\cite{Goncharov:1985ka}
\bibitem{Goncharov:1985ka}
  A.~S.~Goncharov and A.~D.~Linde,
  %``A Simple Realization Of The Inflationary Universe Scenario In SU(1,1)
  %Supergravity,''
  Class.\ Quant.\ Grav.\  {\bf 1}, L75 (1984).
  %%CITATION = CQGRD,1,L75;%%

%\cite{Murayama:1993xu}
\bibitem{Murayama:1993xu}
  H.~Murayama, H.~Suzuki, T.~Yanagida and J.~Yokoyama,
  %``Chaotic inflation and baryogenesis in supergravity,''
  Phys.\ Rev.\  D {\bf 50}, 2356 (1994)
  [arXiv:hep-ph/9311326].
  %%CITATION = PHRVA,D50,2356;%%


%\cite{Kawasaki:2000yn}
\bibitem{Kawasaki:2000yn}
  M.~Kawasaki, M.~Yamaguchi and T.~Yanagida,
  %``Natural chaotic inflation in supergravity,''
  Phys.\ Rev.\ Lett.\  {\bf 85}, 3572 (2000)
  [arXiv:hep-ph/0004243].
  %%CITATION = PRLTA,85,3572;%%

%\cite{Kawasaki:2000ws}
\bibitem{Kawasaki:2000ws}
  M.~Kawasaki, M.~Yamaguchi and T.~Yanagida,
  %``Natural chaotic inflation in supergravity and leptogenesis,''
  Phys.\ Rev.\  D {\bf 63}, 103514 (2001)
  [arXiv:hep-ph/0011104].
  %%CITATION = PHRVA,D63,103514;%%

%\cite{Kadota:2007nc}
\bibitem{Kadota:2007nc}
  K.~Kadota and M.~Yamaguchi,
  %``D-term chaotic inflation in supergravity,''
  Phys.\ Rev.\  D {\bf 76}, 103522 (2007)
  [arXiv:0706.2676 [hep-ph]].
  %%CITATION = PHRVA,D76,103522;%%

%\cite{Kawano:2007gg}
\bibitem{Kawano:2007gg}
  T.~Kawano,
  %``Chaotic D-Term Inflation,''
  arXiv:0712.2351 [hep-th].
  %%CITATION = ARXIV:0712.2351;%%

%\cite{Kadota:2008pm}
\bibitem{Kadota:2008pm}
  K.~Kadota, T.~Kawano and M.~Yamaguchi,
  %``New D-term chaotic inflation in supergravity and leptogenesis,''
  arXiv:0802.0525 [hep-ph].
  %%CITATION = ARXIV:0802.0525;%%

%\cite{Liddle:2006ev}
\bibitem{Liddle:2006ev}
  A.~R.~Liddle, D.~Parkinson, S.~M.~Leach and P.~Mukherjee,
  %``The WMAP normalization of inflationary cosmologies,''
  Phys.\ Rev.\  D {\bf 74}, 083512 (2006)
  [arXiv:astro-ph/0607275].
  %%CITATION = PHRVA,D74,083512;%%

\bibitem{LAMBDA}
{\tt http://lambda.gsfc.nasa.gov/product/map/current/likelihood\_get.cfm}
    
%\cite{Kawasaki:2004qu}
\bibitem{Kawasaki:2004qu}
  M.~Kawasaki, K.~Kohri and T.~Moroi,
  %``Big-bang nucleosynthesis and hadronic decay of long-lived massive
  %particles,''
  Phys.\ Rev.\  D {\bf 71} (2005) 083502
  [arXiv:astro-ph/0408426].
  %%CITATION = PHRVA,D71,083502;%%
  
  %\cite{Kawasaki:2004yh}
\bibitem{Kawasaki:2004yh}
  M.~Kawasaki, K.~Kohri and T.~Moroi,
  %``Hadronic decay of late-decaying particles and big-bang nucleosynthesis,''
  Phys.\ Lett.\  B {\bf 625}, 7 (2005)
  [arXiv:astro-ph/0402490].
  %%CITATION = PHLTA,B625,7;%%

%\cite{Kawasaki:1999na}
\bibitem{Kawasaki:1999na}
   M.~Kawasaki, K.~Kohri and N.~Sugiyama,
   %``Cosmological Constraints on Late-time Entropy Production,''
   Phys.\ Rev.\ Lett.\  {\bf 82}, 4168 (1999)
   [arXiv:astro-ph/9811437].
   %%CITATION = PRLTA,82,4168;%%

%\cite{Kawasaki:2000en}
\bibitem{Kawasaki:2000en}
   M.~Kawasaki, K.~Kohri and N.~Sugiyama,
   %``MeV-scale reheating temperature and thermalization of neutrino
   %background,''
   Phys.\ Rev.\  D {\bf 62}, 023506 (2000)
   [arXiv:astro-ph/0002127].
   %%CITATION = PHRVA,D62,023506;%%

%\cite{Hannestad:2004px}
\bibitem{Hannestad:2004px}
   S.~Hannestad,
   %``What is the lowest possible reheating temperature?,''
   Phys.\ Rev.\  D {\bf 70}, 043506 (2004)
   [arXiv:astro-ph/0403291].
   %%CITATION = PHRVA,D70,043506;%%

%\cite{Ichikawa:2005vw}
\bibitem{Ichikawa:2005vw}
   K.~Ichikawa, M.~Kawasaki and F.~Takahashi,
   %``The oscillation effects on thermalization of the neutrinos in the  universe
   %with low reheating temperature,''
   Phys.\ Rev.\ D {\bf 72}, 043522 (2005)
  [arXiv:astro-ph/0505395].
   %%CITATION = ASTRO-PH 0505395;%%

%\cite{Ichikawa:2006vm}
\bibitem{Ichikawa:2006vm}
   K.~Ichikawa, M.~Kawasaki and F.~Takahashi,
   %``Constraint on the Effective Number of Neutrino Species from the WMAP and
   %SDSS LRG Power Spectra,''
   JCAP {\bf 0705}, 007 (2007)
   [arXiv:astro-ph/0611784].
   %%CITATION = JCAPA,0705,007;%%
  
%\cite{Kumekawa:1994gx}
\bibitem{Kumekawa:1994gx}
  K.~Kumekawa, T.~Moroi and T.~Yanagida,
  %``Flat potential for inflaton with a discrete R invariance in supergravity,''
  Prog.\ Theor.\ Phys.\  {\bf 92}, 437 (1994)
  [arXiv:hep-ph/9405337].
  %%CITATION = PTPKA,92,437;%%

%\cite{Izawa:1996dv}
\bibitem{Izawa:1996dv}
  K.~I.~Izawa and T.~Yanagida,
  %``Natural new inflation in broken supergravity,''
  Phys.\ Lett.\  B {\bf 393}, 331 (1997)
  [arXiv:hep-ph/9608359].
  %%CITATION = PHLTA,B393,331;%%

%\cite{Linde:1991km}
\bibitem{Linde:1991km}
  A.~D.~Linde,
  %``Axions in inflationary cosmology,''
  Phys.\ Lett.\  B {\bf 259}, 38 (1991).
  %%CITATION = PHLTA,B259,38;%%

%\cite{Linde:1993cn}
\bibitem{Linde:1993cn}
  A.~D.~Linde,
  %``Hybrid inflation,''
  Phys.\ Rev.\  D {\bf 49}, 748 (1994)
  [arXiv:astro-ph/9307002].
  %%CITATION = PHRVA,D49,748;%%

%\cite{Copeland:1994vg}
\bibitem{Copeland:1994vg}
  E.~J.~Copeland, A.~R.~Liddle, D.~H.~Lyth, E.~D.~Stewart and D.~Wands,
  %``False vacuum inflation with Einstein gravity,''
  Phys.\ Rev.\  D {\bf 49}, 6410 (1994)
  [arXiv:astro-ph/9401011].
  %%CITATION = PHRVA,D49,6410;%%

%\cite{Dvali:1994ms}
\bibitem{Dvali:1994ms}
  G.~R.~Dvali, Q.~Shafi and R.~K.~Schaefer,
  %``Large scale structure and supersymmetric inflation without fine tuning,''
  Phys.\ Rev.\ Lett.\  {\bf 73}, 1886 (1994)
  [arXiv:hep-ph/9406319].
  %%CITATION = PRLTA,73,1886;%%

%\cite{Lazarides:1996dv}
\bibitem{Lazarides:1996dv}
  G.~Lazarides, R.~K.~Schaefer and Q.~Shafi,
  %``Supersymmetric inflation at the grand unification scale,''
  Phys.\ Rev.\  D {\bf 56}, 1324 (1997)
  [arXiv:hep-ph/9608256].
  %%CITATION = PHRVA,D56,1324;%%

%\cite{Lazarides:2007dg}
\bibitem{Lazarides:2007dg}
  G.~Lazarides and C.~Pallis,
  %``Reducing the spectral index in F-term hybrid inflation through a
  %complementary modular inflation,''
  Phys.\ Lett.\  B {\bf 651}, 216 (2007)
  [arXiv:hep-ph/0702260].
  %%CITATION = PHLTA,B651,216;%%

%\cite{Moroi:2002rd}
\bibitem{Moroi:2002rd}
  T.~Moroi and T.~Takahashi,
  %``Cosmic density perturbations from late-decaying scalar condensations,''
  Phys.\ Rev.\  D {\bf 66}, 063501 (2002)
  [arXiv:hep-ph/0206026].
  %%CITATION = PHRVA,D66,063501;%%

%\cite{Lyth:2002my}
\bibitem{Lyth:2002my}
  D.~H.~Lyth, C.~Ungarelli and D.~Wands,
  %``The primordial density perturbation in the curvaton scenario,''
  Phys.\ Rev.\  D {\bf 67}, 023503 (2003)
  [arXiv:astro-ph/0208055].
  %%CITATION = PHRVA,D67,023503;%%

%\cite{Lyth:2003ip}
\bibitem{Lyth:2003ip}
  D.~H.~Lyth and D.~Wands,
  %``The CDM isocurvature perturbation in the curvaton scenario,''
  Phys.\ Rev.\  D {\bf 68}, 103516 (2003)
  [arXiv:astro-ph/0306500].
  %%CITATION = PHRVA,D68,103516;%%
 
%\cite{Zaldarriaga:2003my}
\bibitem{Zaldarriaga:2003my}
  M.~Zaldarriaga,
  %``Non-Gaussianities in models with a varying inflaton decay rate,''
  Phys.\ Rev.\  D {\bf 69}, 043508 (2004)
  [arXiv:astro-ph/0306006].
  %%CITATION = PHRVA,D69,043508;%%

%\cite{Bartolo:2003bz}
\bibitem{Bartolo:2003bz}
  N.~Bartolo, S.~Matarrese and A.~Riotto,
  %``Evolution of second-order cosmological perturbations and
  %non-Gaussianity,''
  JCAP {\bf 0401}, 003 (2004)
  [arXiv:astro-ph/0309692].
  %%CITATION = JCAPA,0401,003;%%

%\cite{Vernizzi:2003vs}
\bibitem{Vernizzi:2003vs}
  F.~Vernizzi,
  %``Cosmological perturbations from varying masses and couplings,''
  Phys.\ Rev.\  D {\bf 69}, 083526 (2004)
  [arXiv:astro-ph/0311167].
  %%CITATION = PHRVA,D69,083526;%%

%\cite{Jokinen:2005by}
\bibitem{Jokinen:2005by}
  A.~Jokinen and A.~Mazumdar,
  %``Very Large Primordial Non-Gaussianity from multi-field: Application to
  %Massless Preheating,''
  JCAP {\bf 0604}, 003 (2006)
  [arXiv:astro-ph/0512368].
  %%CITATION = JCAPA,0604,003;%%

%\cite{Barnaby:2006km}
\bibitem{Barnaby:2006km}
  N.~Barnaby and J.~M.~Cline,
  %``Nongaussianity from tachyonic preheating in hybrid inflation,''
  Phys.\ Rev.\  D {\bf 75}, 086004 (2007)
  [arXiv:astro-ph/0611750].
  %%CITATION = PHRVA,D75,086004;%%

%\cite{Chambers:2007se}
\bibitem{Chambers:2007se}
  A.~Chambers and A.~Rajantie,
  %``Lattice calculation of non-Gaussianity from preheating,''
  Phys.\ Rev.\ Lett.\  {\bf 100}, 041302 (2008)
  [arXiv:0710.4133 [astro-ph]].
  %%CITATION = PRLTA,100,041302;%%

%\cite{Ichikawa:2008ne}
\bibitem{Ichikawa:2008ne}
  K.~Ichikawa, T.~Suyama, T.~Takahashi and M.~Yamaguchi,
  %``Primordial Curvature Fluctuation and Its Non-Gaussianity in Models with
  %Modulated Reheating,''
  arXiv:0807.3988 [astro-ph].
  %%CITATION = ARXIV:0807.3988;%%


\end{thebibliography}
\end{document}